\documentclass[preprint,prd,aps]{revtex4}
\usepackage{graphicx}
\usepackage{epsfig}
\usepackage{amssymb}
\begin{document}
\title{Quantum-classical correspondence for a dc-biased cavity resonator--Cooper-pair transistor system\footnote{To appear in {\it Fluctuating Nonlinear Oscillators}, Edited by Mark Dykman (Oxford University Press).}}
\author{M.P. Blencowe}\affiliation{Department of Physics and Astronomy, Dartmouth College, Hanover, New Hampshire 03755, USA }
\author{A.D. Armour}\affiliation{School of Physics and Astronomy, University of Nottingham, Nottingham NG7 2RD, United Kingdom}
\author{A.J. Rimberg}\affiliation{Department of Physics and Astronomy, Dartmouth College, Hanover, New Hampshire 03755, USA }

\date{\today}

\begin{abstract}
We investigate the quantum versus classical dynamics of a microwave cavity-coupled-Cooper pair transistor (CPT) system, where  an  applied dc bias causes the system to self-oscillate via the ac Josephson effect. Varying the dc bias allows the self-oscillation frequency to be tuned.  An unusual feature of the system design is that the dc bias does not significantly affect the high quality factor  of the cavity mode to which the CPT predominantly couples. The  CPT-cavity mode system has a mechanical analogue involving a driven coupled pendulum-oscillator system. The  corresponding, nonlinear classical dynamical equations exhibit chaotic, as well as aperiodic motions depending on the initial conditions and the nature and strengths of the damping/noise forces. The quantum master equation exhibits such phenomena as dynamical tunnelling and the generation of nonclassical states from initial classical states.  Obviating the need for an external ac-drive line, which typically is harder to noise filter than a dc bias line,  the self-oscillating  system described here has considerable promise for  demonstrating macroscopic quantum dynamical behavior.
\end{abstract}

\maketitle

\section{Introduction}
The work presented in this chapter has its origins in a seemingly mundane microwave engineering question: is it possible to apply a dc voltage (or current) bias to the center conductor of a superconducting coplanar microwave cavity, without significantly affecting the quality factor of, say, the first and second microwave modes of the cavity? Our original motivation behind this question was to devise a circuit quantum electrodynamics (QED) based scheme~\cite{wallraffnature04} that can generate and detect quantum states of a mechanical resonator~\cite{armournjp08,blencowenjp08}, where the dc bias is required to strongly couple a nanomechanical resonator to a superconducting qubit. However, it turns out that having such a dc bias functionality opens up possibilities for other heretofore difficult-to-realize quantum dynamical investigations, one of which we shall focus on here.

We shall in particular investigate  the quantum dynamics of the device shown in Fig.~\ref{cavityfig}, which comprises two Josephson junctions (JJ) in series with a gate electrode, and where the source electrode to one of the JJ's contacts the center conductor of the microwave cavity, while the drain electrode from the other JJ contacts the ground plane of the microwave cavity. The following section describes how the microwave cavity design allows the application of a dc voltage bias $V_{\mathrm{dc}}$ to the center conductor, while maintaining a very large quality factor of the second microwave mode to which the JJ's strongly couple~\cite{chenapl11}. For  not too large a $V_{\mathrm{dc}}$ bias, the JJ's operate in the subgap region as a ``Cooper pair transistor" (CPT), where the dc bias generates a tunable oscillating supercurrent through the CPT via the ac Josephson effect. The tunneling Cooper pairs will both emit into and absorb photons from the second microwave mode, and it is the resulting coupled CPT-cavity mode quantum dynamics  that will be of central interest to us.

Related devices comprising one or more JJ's embedded in a microwave cavity date back to just a few years following the discovery of the ac Josephson effect~\cite{josephsonpl62}, where classical signatures of the resonant microwave modes of the tunnel junctions themselves, interacting with the alternating tunnel currents, were observed and discussed~\cite{werthamer,zimmerman,smith}. Beginning in the `90's, investigations addressed the effect of a  structured electromagnetic environment with resonant modes on the current-voltage characteristics of dc voltage biased JJ's~\cite{holstprl94,ingoldprb94,hofheinzprl01}. And more recently, similar investigations involving double JJ devices were carried out~\cite{leppakangasprb08,pashkinprb11}. However, the quality factors of the electromagnetic modes in these devices were small, typically less than 10, to be contrasted with quality factors exceeding $10^3$ for the present device design~\cite{chenapl11} shown in Fig.~\ref{cavityfig}. As a consequence,  emitted microwave photons will now remain in the cavity mode for many Cooper pair tunnel oscillation cycles before leaking out of the cavity mode; it does not make sense to treat the microwave cavity as an electromagnetic environment for the CPT. Instead, the cavity and CPT should be viewed as a strongly-coupled, quantum coherent system.

The  CPT-cavity mode device has a mechanical analogue involving a driven coupled pendulum-oscillator system (Sec.~\ref{mechanicalsec}). The  corresponding, nonlinear classical dynamical equations exhibit chaotic, as well as aperiodic motions depending on the initial conditions and the nature and strengths of the damping/noise forces. Thus, the device in principle allows the experimental investigation of the quantum dynamics of a system for which the corresponding classical dynamics is chaotic. There is a long tradition of using Josephson junction devices for investigating macroscopic quantum dynamics in systems with corresponding nonlinear classical equations~\cite{srivastavapr87,leggettcp09}. The Sussex group carried out some of the first, pioneering work in the `80's~\cite{prancehpa83}, which was followed by the  demonstration of quantum tunneling  by the Clarke group at Berkeley~\cite{clarkescience88}, and which culminated in  demonstrations over a decade later of superposition states by the Lukens~\cite{friedmannature00} and Mooij~\cite{chiorescuscience03} groups at Stonybrook and Delft, respectively. Subsequent, related developments have largely focussed on the realization of superconducting quantum bits for quantum computing applications~\cite{neeleynature10,dicarlonature10}, although JJ devices still occasionally are used for exploring macroscopic quantum dynamics and the transition to classical dynamics~\cite{fedorovprl11}.

A large body of theoretical work concerning the quantum-classical correspondence for driven  systems has focused on the Duffing and  other anharmonic oscillators~\cite{dykmanzetf88,zurekprl94,brunjpa96,kohlerpre97,habibprl98,bhattacharyaprl00,monteolivapre01,habibprl02,peanoprb04,everittpre05,marthalerpra06,dykmanpre07,greenbaumpre07,serbanprl07,katznjp08,versopra10,ketzmerickpre10}, as well as on various rigid rotor models~\cite{todaptps89,casatiptps89,foxpra90,foxpra91,grahampra91,grahamprl91,foxpre94,latkapra94,mirbachprl95,gorincp97,mouchetpre01,mouchetpre06,reichl04,haake10}; many insights have been gained by investigating dynamical properties using a quantum phase space  (i.e., Wigner or Husimi function~\cite{takahashiptps89,leepre93}) description, and by examining the Floquet states and associated quasienergy spectra. However, relevant experimental results have been few~\cite{steckscience01,chaudhurynature09,vijayrsi09}. One of the key difficulties is that most experimental realisations require an {\it external} ac signal to drive the system, which can be one of the most significant sources of noise, preventing the system from displaying manifest quantum dynamical behavior. In contrast,  the CPT-resonator system in Fig.~\ref{cavityfig} generates its own ac drive, i.e., it self-oscillates. As a consequence of the ac-Josephson effect,  only a dc voltage bias $V_{\mathrm{dc}}$ is required, and by varying $V_{\mathrm{dc}}$, the drive frequency can be tuned.   Since it is considerably easier to noise filter  a dc bias line than an ac-drive line, the  device described here has considerable promise for exhibiting macroscopic quantum dynamical behavior.

The outline of this chapter is as follows. In Sec.~\ref{devicesec}, we give a description of the CPT-cavity system. The classical system equations are derived in Sec.~\ref{classicalsec}, and solutions to these equations are discussed in Sec.~\ref{sec:classd}. The corresponding quantum master equation is derived and a quantum phase space representation of the system state is described in Sec.~\ref{sec:quantumeq}. Solutions to the quantum master equation are discussed in Sec.~\ref{quantumdynsec} and in Sec.~\ref{classicallimitsec} we investigate the classical limit of the quantum master equation. We conclude in Sec.~\ref{sec:conclusion}.

Much of the analysis will in fact deal with a simplified system comprising the driven `pendulum' part  of the device. An analysis of the full CPT-resonator mode system dynamics, including  results from experiment, will be published elsewhere.

\section{The cavity-Cooper pair transistor device}
\label{devicesec}

\begin{figure}
\begin{center}
\includegraphics[width=9cm]{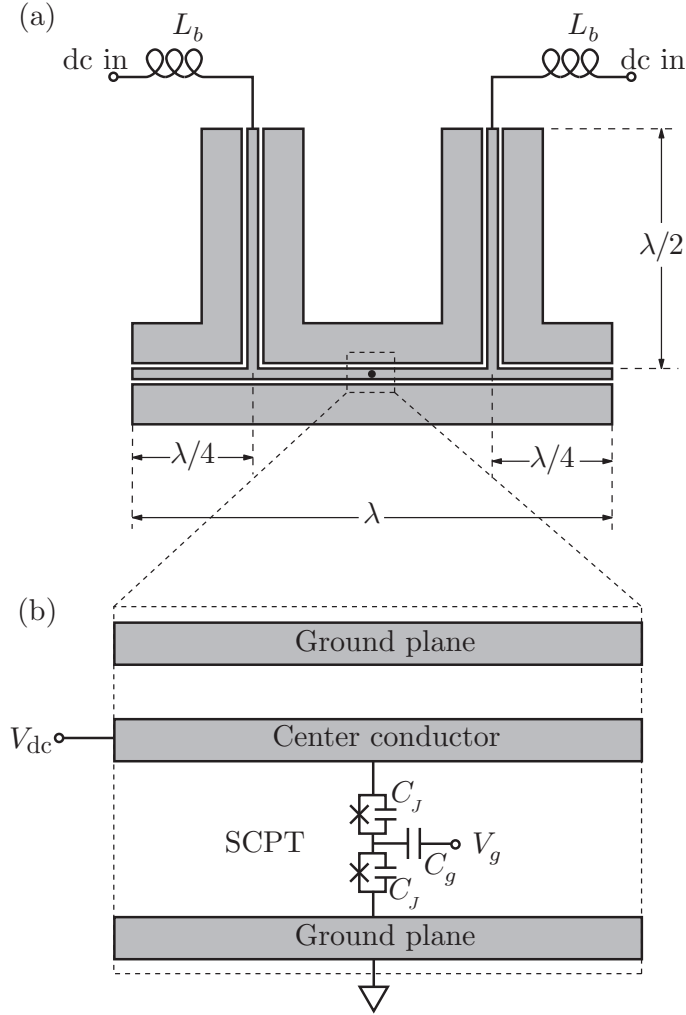} 
\caption{\label{cavityfig} (a) Schematic diagram of a dc biased microwave cavity, showing the location of the inductively terminated bias lines and the sample location (black dot). (b) Illustration of a CPT embedded in a dc-biased cavity at the central voltage antinode.  }
\end{center}
\end{figure}

To introduce a dc bias into a high-$Q$ microwave cavity, we begin with a standard coplanar-waveguide-based resonator  that is one wavelength $\lambda$ long at the operating frequency,  illustrated schematically in Fig.~\ref{cavityfig}(a).  As is usually the case, the ends of the cavity are terminated by small capacitors (on the order of a few fF) that even at a typical operating frequency of 5~GHz have a large impedance. To a first approximation, then, we can treat these terminations as open circuits, so that  the cavity voltage is a local maximum at the cavity end, and the cavity current a local minimum.  At a distance $\lambda/4$ from each end of the cavity, the situation is reversed: the cavity voltage is minimal and the current maximal, so that the $\lambda/4$ points are low impedance points.

At the $\lambda/4$ low impedance points we then introduce dc bias lines consisting of sections of waveguide terminated with an inductance $L_{b}$.  These lines are chosen to have a length $\lambda/2$, so that the impedance they present to the main cavity line at the $\lambda/4$ point is the same as their terminating impedance $i \omega L_{b}$.  For even a small inductance of a few nH, this impedance can be substantial at the operating frequency of the cavity.

A microwave photon approaching either dc biasing ``T'' junction will therefore see a short circuit (the low impedance of the main line) in parallel with a large impedance (the dc bias line) and to first order the cavity photons will be unaffected by the presence of the dc bias lines.  The second (full wave) resonance of the cavity should still enjoy a very large $Q$ of up to several thousand in the presence of a dc bias.   By placing a CPT at the center of the cavity (the black dot in Fig.~\ref{cavityfig}(a)), where there is an antinode in the cavity voltage, it should be possible to strongly couple the CPT to the cavity and use an applied dc bias to produce self oscillations of the CPT/cavity system via the ac Josephson effect.

\begin{figure}
\begin{center}
\includegraphics[width=9cm]{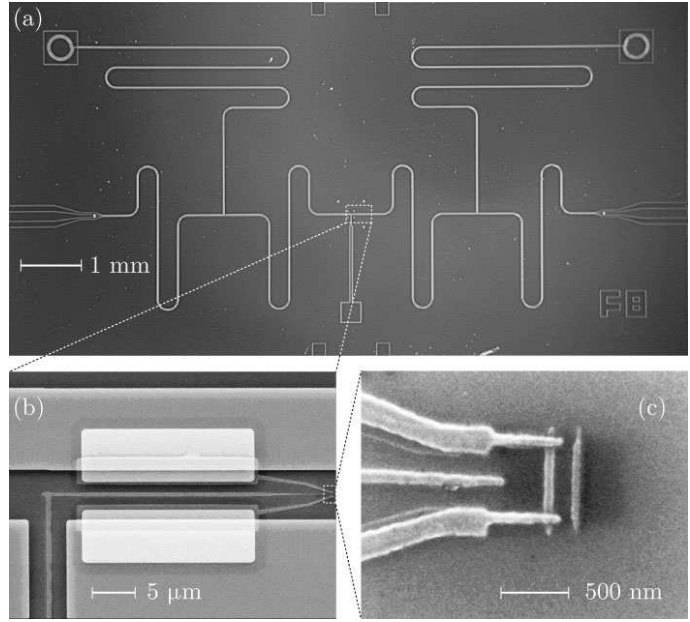} 
\caption{\label{imagefig} (a)  Optical micrograph of a microwave cavity with inductively terminated dc bias lines.  A contact pad for the CPT gate is visible at the bottom center. (b) Electron micrograph of the center of the cavity showing the gate lead entering the cavity at lower left, the bright Ti/Au contact pads, and a CPT at the right. (c) Detailed view of the CPT\@.  The gate lead is at left the the Josephson junctions at the top and bottom of the central island.  }
\end{center}
\end{figure}

Electron and optical micrographs of a device based on these ideas are shown in Fig.~\ref{imagefig}.  The cavity itself is fabricated out of a Nb film on an undoped Si substrate, as shown in Fig.~\ref{imagefig}(a).  Input and output lines on the left and right are coupled to the main line by small capacitors.  The dc bias lines extend toward the top of the image; each is terminated by a small on-chip spiral inductor.   Cavities based on this design have been shown to posses a large $Q$ of several thousand for the full wave mode at a temperature of 4~K even when a dc bias voltage or current is applied to the central conductor of the cavity \cite{chenapl11}.

At the center of the cavity, a narrow wire to be used as a gate line for the CPT is brought through the ground plane of the waveguide, as shown in Fig.~\ref{imagefig}(b).  Two thin Ti/Au contact pads are added to the central conductor and ground plane of the cavity to the right of the entry point for the gate wire.  These contact pads, which are driven superconducting by the proximity effect, allow for good metal-to-metal contact between the CPT and cavity.  Finally, the CPT and its gate are added to the structure using standard electron beam lithography and shadow evaporation techniques, as in Fig.~\ref{imagefig}(c).

\begin{figure}
\center{
\includegraphics[height=2.5in]{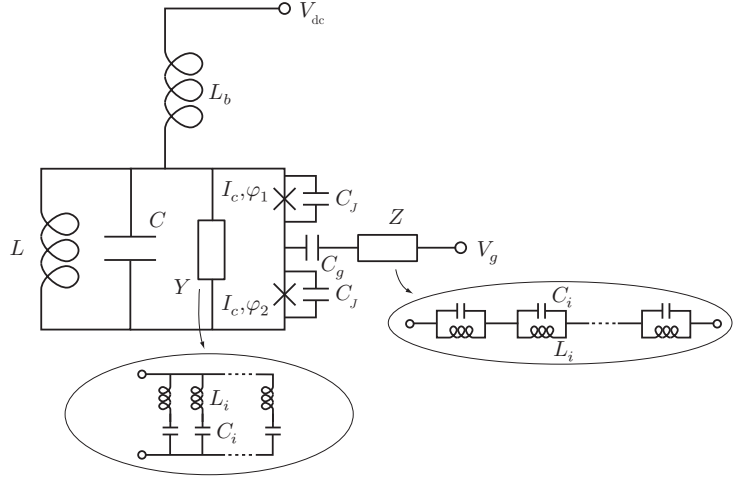}}
 \caption{Effective circuit model of the cavity-CPT system. The cavity is modeled as a lumped element circuit consisting of the capacitance $C$ and inductance $L$, which are in parallel with the CPT that is formed by two Josephson junctions in series. The circuit is controlled by the two voltages $V_g$ and $V_{dc}$. Dissipation is included by incorporating an effective admittance, $Y(\omega)$, in parallel with $C$ and $L$ and an impedance $Z(\omega)$ in series with the gate voltage.}\label{fig:circuit}
\end{figure}

\section{Classical model of device}
\label{classicalsec}
\subsection{Closed system equations}
The effective lumped element model description of the dc voltage $V_{\mathrm{dc}}$ biased microwave cavity-coupled Cooper pair transistor (CPT) device is illustrated in Fig.~\ref{fig:circuit}. It is supposed that, for the considered $V_{\mathrm{dc}}$ bias range, the CPT couples predominantly to a particular mode of the cavity. We neglect for the time being the cavity and CPT sources of dissipation, modeled by the parallel $LC$ network admittance and series $LC$ network impedance, respectively, focusing first on writing down the closed system equations of motion. For a typical device, the cavity effective capacitance $C$ is a few pF, while the Josephson junction (JJ) capacitance $C_J$ is at least a few hundred aF, and the gate bias capacitance is about $10~{\mathrm{aF}}$. Furthermore, the effective bias line inductance $L_b$ is a few nH and the cavity effective inductance $L$ is a few tenths of nH. Thus, the typical size hierarchies are $L_{b} \gg L$ and $C \gg C_{J} \gg C_{g}$. We shall make use of these to simplify by approximation the equations of motion. Using Kirchhoff's Laws  and the constitutive relations for the various lumped circuit elements, it is straightforward to obtain the equations of motion. In terms of the phase differences across the two JJs, the equations are
\begin{eqnarray}
 (C+C_J)\ddot{\varphi}_1 &+&C\ddot{\varphi}_2+{L}^{-1}(\varphi_1+\varphi_2-\varphi^0_1-\varphi^0_2)=\cr
&-&\frac{2\pi I_c}{\Phi_0}\sin\varphi_1+\frac{2\pi}{\Phi_0}L_b^{-1}V_{\mathrm{dc}}t
 \label{phi1eq}
 \end{eqnarray}
and
\begin{eqnarray}
 (C+C_J+C_g)\ddot{\varphi}_2 &+&C\ddot{\varphi}_1+{L}^{-1}(\varphi_1+\varphi_2-\varphi^0_1-\varphi^0_2)=\cr
 &-&\frac{2\pi I_c}{\Phi_0}\sin\varphi_2
 +\frac{2\pi}{\Phi_0}L_b^{-1}V_{\mathrm{dc}}t+\frac{2\pi}{\Phi_0}C_g\dot{V}_g,
 \label{phi2eq}
 \end{eqnarray}
where $\Phi_0=h/(2e)$ is the flux quantum, $I_c$ is the JJ critical current, $\varphi_i^0$ is an integration constant, and we have assumed $L\ll L_b$. Transforming to `center-of-mass' (CoM) and relative phase coordinates $\gamma_{\pm}=(\varphi_1\pm\varphi_2)/2$, Eqs.~(\ref{phi1eq}) and (\ref{phi2eq}) become
 \begin{equation}
 C\ddot{\gamma}_+ +{L}^{-1}(\gamma_+-\gamma_+^0) =-\frac{\pi I_c}{\Phi_0}\sin\gamma_+\cos\gamma_-  +\frac{\pi}{\Phi_0}L_b^{-1}V_{\mathrm{dc}}t
 \label{gamma+eq}
 \end{equation}
 and
  \begin{equation}
 C_J\ddot{\gamma}_- =-\frac{2\pi I_c}{\Phi_0}\sin\gamma_-\cos\gamma_+ -\frac{\pi}{\Phi_0}C_g\dot{V}_g,
 \label{gamma-eq}
 \end{equation}
where we have assumed $ C_g\ll C_J\ll C$.

Eqs.~(\ref{gamma+eq}) and (\ref{gamma-eq}) follow via the Euler-Lagrange equations from the Lagrangian
\begin{eqnarray}
{\mathcal{L}}&=&\frac{1}{2}\left(\frac{\Phi_0}{\pi}\right)^2 C \dot{\gamma}_+^2 +\frac{1}{4}\left(\frac{\Phi_0}{\pi}\right)^2 C_J\dot{\gamma}_-^2 +\frac{\Phi_0}{\pi}L_b^{-1}V_{\mathrm{dc}} t \gamma_+ + \frac{\Phi_0}{2\pi}C_g V_g \dot{\gamma}_-\cr
&& -\frac{1}{2} \left(\frac{\Phi_0}{\pi}\right)^2 {L}^{-1}(\gamma_+-\gamma_+^0)^2+\frac{\Phi_0}{\pi} I_c \cos\gamma_-\cos\gamma_+.
\label{lagrangianeq}
\end{eqnarray}
The Hamiltonian is
\begin{eqnarray}
{\mathcal{H}}&=&\left(\frac{\pi}{\Phi_0}\right)^2(2 C)^{-1}p_+^2+\left(\frac{\Phi_0}{\pi}\right)^2 (2L)^{-1} (\gamma_+-\gamma_+^0)^2 -\frac{\Phi_0}{\pi} L_b^{-1} V_{\mathrm{dc}} t\gamma_+\cr
&& +E_{C_J} (N-N_g)^2 -2 E_J \cos\gamma_+\cos\gamma_-,
\label{hamiltonianeq}
\end{eqnarray}
where $N=p_-/\hbar$ is minus the number of excess Cooper pairs on the island, $N_g=C_g V_g/(2e)$ is the polarization charge induced by the applied gate voltage bias $V_g$ in units of Cooper pair charge, $E_{C_J}=(2e)^2/(2\cdot2C_J)=e^2/C_J$ is the approximate CPT charging energy (neglecting $C_g$), i.e., the electrostatic energy cost for putting one additional Cooper pair on the CPT island, and $E_J =I_c\Phi_0/(2\pi)$ is the Josephson energy of a single JJ.

It is convenient to work instead in terms of the shifted CoM coordinate: $\tilde{\gamma}_+=\gamma_+ -\gamma_+^0-\omega_d t$, where the driving frequency is
\begin{equation}
\omega_d=\frac{L}{L_b}\frac{eV_{\mathrm{dc}}}{\hbar}.
\label{drivefreq2eq}
\end{equation}
Performing this canonical transformation with the appropriate generating function, we obtain the following transformed Hamiltonian:
\begin{eqnarray}
{\mathcal{H}}&=&\left(\frac{\pi}{\Phi_0}\right)^2(2 C)^{-1}p_+^2+\left(\frac{\Phi_0}{\pi}\right)^2 (2L)^{-1}\gamma_+^2 +E_{C_J} (N-N_g)^2 \cr
&& -2 E_J\cos\gamma_- \cos(\gamma_+ +\omega_d t),
\label{transformedhamiltonian3eq}
\end{eqnarray}
where we have dropped the tilde on the shifted $\gamma_+$ coordinate and have set $\gamma_+^0 =0$.

The key observation to make about  Hamiltonian~(\ref{transformedhamiltonian3eq}) is the presence of the time-dependent drive, which originates from the ac Josephson effect, and can be controlled via the externally applied $V_{\mathrm{dc}}$ bias [Eq.~(\ref{drivefreq2eq})]; the nonlinear system self-oscillates. In contrast to most other driven nonlinear system investigations, no externally applied ac drive is required, thus eliminating one of the main sources of noise that hinders the demonstration of macroscopic quantum dynamics.

\subsection{Mechanical analogue model}
\label{mechanicalsec}

In order to gain insights into the cavity mode-CPT dynamics, as well as motivate other parameter choices, it is useful to consider a mechanical analogue. Hamiltonian~(\ref{transformedhamiltonian3eq}) can be reexpressed in the following  form:
\begin{equation}
{\mathcal{H}}=\frac{J_+^2}{2I_+}+\frac{1}{2}I_+\omega_+^2 \gamma_+^2 +\frac{1}{2 I_-}\left[J_- -\left(\frac{\Phi_0}{2\pi}\right)C_g V_g\right]^2-I_- \omega_-^2 \cos(\gamma_+ +\omega_d t)\cos\gamma_- ,
\label{mechanicalhamiltonianeq}
\end{equation}
where  $I_+ =C(\Phi_0/\pi)^2$, $I_-=\hbar^2/(2 E_{C_J})$, $\omega_+=1/\sqrt{LC}$, and $\omega_-=2\sqrt{E_J E_{C_J}}/\hbar$.  From Eq.~(\ref{mechanicalhamiltonianeq}), we see that the cavity mode-CPT system is equivalent to a system consisting of two coupled rotors with moments of inertia $I_{\pm}$ and angular momentum $J_{\pm}$.  Neglecting the rotor coupling, the `+' rotor behaves as a torsional oscillator with frequency $\omega_+$. For small $\gamma_+$ angular displacements and with the drive turned off (i.e., $V_{\mathrm{dc}}=0$), the `-' rotor behaves as a pendulum. For small $\gamma_-$ displacements, the pendulum oscillates approximately harmonically with frequency $\omega_-$. With the drive turned on (i.e., $V_{\mathrm{dc}}\neq 0$), the pendulum rotor's `gravitational acceleration' is sinusoidally modulated at frequency $\omega_d$, periodically switching sign as a result. The gravitational acceleration is also modulated by the torsional oscillator coordinate. The ratio of the rotors' moments of inertia is
\begin{equation}
\frac{I_+}{I_-}=\frac{R_K}{Z}\frac{E_{C_J}}{\hbar\omega_+}=\frac{2C}{C_J},
\label{massratioeq}
\end{equation}
where $Z=\pi\sqrt{L/C}\approx 50~\Omega$ is the cavity impedance and $R_K=h/e^2 \approx 25.8~{\mathrm{k}\Omega}$ is the von Klitzing constant. The frequency ratio for small angle, undriven oscillations is
\begin{equation}
\frac{\omega_-}{\omega_+}=2\frac{\sqrt{E_J E_{C_J}}}{\hbar\omega_+}.
\label{frequencyratioeq}
\end{equation}
For typical capacitance values, CPT charging and Josephson energies of a few Kelvins (in units of $k_B$), and for a cavity mode frequency $\omega_+ = 2\pi \times 5~{\mathrm{GHz}}$ ($\hbar\omega_+/k_B =0.24~{\mathrm{K}}$), we see that the ratios in Eqs.~(\ref{massratioeq}) and (\ref{frequencyratioeq}) are large: the moment of inertia ratio is of order $10^4$ and the frequency ratio of  order $10$. Thus, the mechanical analogue corresponds to a  fast pendulum with a small moment of inertia  that is coupled to a slow torsional oscillator with a large moment of inertia.

A measure of the zeropoint fluctuations in the pendulum angular coordinate $\gamma_-$ is
\begin{equation}
\Delta_{zp}^-=\sqrt{\frac{\hbar}{2 I_- \omega_-} }=\sqrt[4]{\frac{4 E_{C_J}}{E_J}}.
\label{gammamzpeq}
\end{equation}
For typical CPT parameter values, we have $\Delta^-_{zp}\approx1$, i.e., the zeropoint uncertainty is comparable to the size of the  $\gamma_-$ coordinate space ($=2\pi$ radians).  Thus, we don't expect the driven quantum pendulum dynamics to resemble much the dynamics of the driven classical pendulum, which can be chaotic. Recovering the classical pendulum limit requires a smaller charging energy than Josephson energy, for example a `transmon'-like CPT~\cite{kochpra07}. The classical limit will be discussed in detail in Sec.~\ref{classicallimitsec}.

How do the mechanical analogue moments of inertia compare in magnitude to those of actual mechanical systems? The hydrogen molecule has a rotational moment of inertia $\approx 5\times 10^{-48}~{\mathrm{kg m^2}}$~\cite{horizfp27}. For  $E_{C_J}\sim 5~{\mathrm{K}}$ ($\equiv 431~\mu{\mathrm{eV}}\equiv 6.9\times 10^{-23}~{\mathrm{J}}$), we have $I_-\approx 8\times 10^{-47}~{\mathrm{kg m^2}}$. Thus, the typical CPT pendulum equivalent  moment of inertia is an order of magnitude larger than that of the hydrogen molecule. For a transmon-like CPT, the moment of inertia is about two orders of magnitude larger than that of the hydrogen molecule. From Eq.~(\ref{massratioeq}), we see that  the cavity mode torsional oscillator equivalent moment of inertia $I_+$ is about $10^5$ times larger than that  of the hydrogen molecule.

\subsection{Open system equations}
The cavity-CPT device is subject to several sources of dissipation and noise. Two significant electromagnetic environment sources arise from the capacitive couplings between cavity and input/output microwave lines and the capacitive coupling between the CPT island and gate voltage bias line. Referring to Fig.\ \ref{fig:circuit} , we model the cavity noise/dissipation by an infinite parallel network of $LC$ `bath' oscillators, and the gate voltage noise/dissipation by an infinite series network of  $LC$ `bath' oscillators~\cite{yurke83,devoret95,burkard04}. The actual dissipative mechanisms can be modeled by such infinite oscillator networks by making appropriate choices for the oscillator frequency distribution spectra. In the following, we will analyze the two noise/dissipation sources independently, beginning first with the cavity noise source.

Extending Hamiltonian~(\ref{transformedhamiltonian3eq}) to include the infinite parallel network of  $LC$ oscillators, we obtain:
\begin{eqnarray}
{\mathcal{H}}&=&\left(\frac{\pi}{\Phi_0}\right)^2(2 C)^{-1}p_+^2+\left(\frac{\Phi_0}{\pi}\right)^2 (2L)^{-1}\gamma_+^2 +E_{C_J} (N-N_g)^2 \cr
&& -2 E_J\cos\gamma_- \cos(\gamma_+ +\omega_d t)\cr
&&+\left(\frac{2\pi}{\Phi_0}\right)^2\sum_i \frac{p_i^2}{2C_i}+\left(\frac{\Phi_0}{2\pi}\right)^2\sum_i \frac{1}{2L_i}(\phi_i-2\gamma_+)^2,
\label{transformeddisshamiltonian2eq}
\end{eqnarray}
where $\phi_i$ is the phase coordinate across the network capacitance $C_i$. Integrating Hamilton's equations of motion for the network oscillator coordinate $\phi_i$, we obtain:
\begin{equation}
\phi_i (t)=\phi_i (0) \cos\omega_i t +\frac{p_i(0)}{m_i\omega_i}\sin\omega_i t +\lambda_i \int_0^t dt'\frac{\sin \omega_i(t-t')}{m_i\omega_i}\gamma_+(t'),
\label{phiisolneq}
\end{equation}
where  $\omega_i=1/\sqrt{L_i C_i}$, the network oscillator ``masses" are $m_i =C_i \left({\Phi_0}/{(2\pi)}\right)^2$ and the system-network oscillator couplings are $\lambda_i=2\left({\Phi_0}/{(2\pi)}\right)^2 {(L_i)^{-1}}$.
Following the approach of Ref.~\cite{cortes85}, we integrate (\ref{phiisolneq}) by parts and substitute into the equations  for $\gamma_+$ and $p_+$ to obtain the following Langevin equation:
\begin{equation}
\ddot{\gamma}_+=-\frac{1}{LC}\gamma_+ -\left(\frac{\pi}{\Phi_0}\right)^2 \frac{ 2 E_J}{C} \cos\gamma_-\sin(\gamma_+ +\omega_d t)-\int_0^t dt' \Gamma(t-t')\dot{\gamma}_+ (t') +f_n(t),
\label{langevineq}
\end{equation}
where we have assumed that the couplings $\lambda_i$ are small and we have neglected frequency renormalization terms and where
\begin{equation}
\Gamma(t) =\left(\frac{\pi}{\Phi_0}\right)^2 \frac{1}{C} \sum_i \frac{\lambda^2_i}{m_i\omega_i^2} \cos\omega_i t=\frac{1}{C}\sum_i\frac{1}{L_i}\cos\omega_i t
\label{kerneleq}
\end{equation}
is the damping kernel
and
\begin{equation}
f_n(t) =\left(\frac{\pi}{\Phi_0}\right)^2 \frac{1}{C} \sum_i \lambda_i \left(\phi_i (0) \cos\omega_i t +\frac{p_i(0)}{m_i\omega_i}\sin\omega_i t\right)
\label{forcenoiseeq}
\end{equation}
is the noise force.
Assuming the network oscillator initial coordinates $\phi_i(0)$, $p_i(0)$ are randomly distributed according to the Maxwell-Boltzmann thermal distribution at temperature $T$, we find for the force-force correlation function:
\begin{equation}
\langle f_n (t) f_n (0)\rangle=\left(\frac{\pi}{\Phi_0}\right)^2 \frac{1}{C} k_B T\ \Gamma (t).
\label{correlationeq}
\end{equation}
With $(\Phi_0/\pi)^2 C$ being  the ``mass" of the  $\gamma_+$ coordinate, we see that (\ref{correlationeq}) obeys the usual fluctuation-dissipation relation.
With the Markovian approximation  $\Gamma(t)\approx \frac{2}{RC} \delta (t)$, Eq.~(\ref{langevineq}) describes a dissipative cavity mode where the admittance in Fig.\ \ref{fig:circuit} is simply replaced by a resistance $R$.

Moving on now to modelling the gate voltage noise, we  insert an infinite series $LC$ network between the gate voltage source $V_g$ and the gate capacitance $C_g$. Hamiltonian~(\ref{transformedhamiltonian3eq}) is then modified approximately as follows:
\begin{eqnarray}
{\mathcal{H}}&=&\left(\frac{\pi}{\Phi_0}\right)^2(2 C)^{-1}\left(p_+ -\sum_i\frac{C_g}{C_i}p_i\right)^2+\left(\frac{\Phi_0}{\pi}\right)^2 (2L)^{-1}\gamma_+^2\cr
&&+E_{C_J} \left(N-N_g+\frac{1}{\hbar}\sum_i\frac{C_g}{C_i}p_i\right)^2 -2 E_J\cos\gamma_- \cos(\gamma_+ +\omega_d t)\cr
&&+\left(\frac{2\pi}{\Phi_0}\right)^2\sum_i \frac{p_i^2}{2C_i}+\left(\frac{\Phi_0}{2\pi}\right)^2\sum_i \frac{1}{2L_i}\phi_i^2,
\label{transformedgatedisshamiltonian2eq}
\end{eqnarray}
where $\phi_i$ is the phase coordinate across the $C_i$ network capacitance, and we assume $C_g$ is small compared to the other capacitances. In the following, we neglect the coupling between the infinite series network $p_i$ and the $p_+$ coordinates, since this results simply in adding to the dissipation due to the cavity mode loss  considered above. Integrating Hamilton's equations of motion for the network oscillator coordinate $p_i$, we obtain:
\begin{eqnarray}
p_i(t)&=&-m_i\omega_i \phi_i(0) \sin\omega_i t +m_i\dot{\phi}_i (0)\cos\omega_i t \cr
&&-m_i\omega_i \lambda_i\int_0^t dt' \sin[\omega_i (t-t')] (N(t')-N_g),
\label{pisolneq}
\end{eqnarray}
where the network oscillator frequencies and masses are the same as for the cavity mode, while the system-network oscillator couplings are now $\lambda_i=(2 E_{C_J}/\hbar) (C_g/C_i)$.
Substituting Eq.~(\ref{pisolneq}) into  Hamilton's equation for $\dot{\gamma}_-$, integrating by parts and neglecting renormalization and shift terms, we obtain the following equation:
\begin{eqnarray}
\dot{\gamma}_-&=&\frac{2 E_{C_J}}{\hbar} (N-N_g)+\frac{1}{\hbar}\sum_i m_i\lambda_i^2 \int_0^t d t' \cos\omega_i (t-t') \dot{N}(t')\cr
&&+\frac{1}{\hbar}\sum_i \lambda_ip_i^{(0)}\cr
&=&\frac{2 E_{C_J}}{\hbar} (N-N_g)-\frac{2 E_J}{\hbar^2}\sum_i m_i\lambda_i^2 \int_0^t d t' \cos\omega_i (t-t')\cr
&&\times \sin\gamma_-(t') \cos(\gamma_+(t')+\omega_d t')+\frac{1}{\hbar}\sum_i \lambda_ip_i^{(0)},
\label{gammadampeq}
\end{eqnarray}
where
\begin{equation}
p_i^{(0)}(t)=-m_i \omega_i \phi_i (0) \sin\omega_i t +m_i \dot{\phi}_i (0)\cos\omega_i t.
\label{bathmomentumeq}
\end{equation}
Assuming the network oscillator initial coordinates $\phi_i(0)$, $\dot{\phi}_i (0)$ are randomly distributed according to the Maxwell-Boltzmann thermal distribution at temperature $T$, we find for the correlation relation:
\begin{eqnarray}
&&\left\langle \left(\sum_i\lambda_i p_i^{(0)}(t)\right)  \left(\sum_i\lambda_i p_i^{(0)}(0)\right)\right\rangle=k_B T \sum_i m_i\lambda_i^2 \cos\omega_i t\cr &&=k_B T \left(\frac{e C_g}{C_J}\right)^2 \sum_i \frac{1}{C_i}\cos\omega_i t.
\label{networkcorreleq}
\end{eqnarray}
Now, we have:
\begin{equation}
\sum_i\lambda_i p^{(0)}_i=\frac{e C_g}{C_J} \frac{\Phi_0}{2\pi}\sum_i\dot{\phi}^{(0)}_i=\frac{e C_g}{C_J} V^{(0)}_{\mathrm{network}},
\label{gateforcenoiseeq}
\end{equation}
where $V^{(0)}_{\mathrm{network}}$ is the fluctuating voltage across the unloaded series network. But in the Markovian approximation, the voltage noise across a resistance is
\begin{equation}
\left\langle V^{(0)}_{\mathrm{network}}(t)V^{(0)}_{\mathrm{network}}(0)\right\rangle =2 k_B T R \delta(t),
\label{networkcorrel2eq}
\end{equation}
and thus
\begin{equation}
\left\langle \left(\sum_i\lambda_i p_i^{(0)}(t)\right)  \left(\sum_i\lambda_i p_i^{(0)}(0)\right)\right\rangle =2 k_B T\left(\frac{e C_g}{C_J}\right)^2 R \delta(t),
\label{gateforcecorreleq}
\end{equation}
with
\begin{equation}
\sum_i \frac{1}{C_i}\cos\omega_i t=2R\delta(t),
\label{markovapproxeq}
\end{equation}
where $R$ is the effective resistance characterizing the loss associated with the gate voltage noise.
Substituting Eq.~(\ref{markovapproxeq}) into the damping term of Eq.~(\ref{gammadampeq}), we obtain for the $\gamma_-$, $N$ coordinate equations in the presence of gate voltage noise and associated damping within the Markov approximation:
\begin{equation}
\dot{\gamma}_-=\frac{2 E_{C_J}}{\hbar} (N-N_g)-\frac{2 E_J}{\hbar^2}\left(\frac{e C_g}{C_J}\right)^2 R\sin\gamma_- \cos(\gamma_+ +\omega_d t) +\frac{1}{\hbar}\sum_i \lambda_i p_i^{(0)}
\label{markovgammaeq}
\end{equation}
and
\begin{equation}
\dot{N}=-\frac{2 E_J}{\hbar} \sin\gamma_-\cos (\gamma_++\omega_d t).
\label{Ndoteq}
\end{equation}

Now that we have analyzed both the cavity noise and gate voltage noise, we finally write down in dimensionless form the classical Markovian Langevin equations for the cavity-CPT system in the presence of both noise sources. In first order form, the  equations of motion are:
\begin{eqnarray}
\dot{\gamma}_+&=&p_+\cr
\dot{p}_+&=&-\gamma_+  +f \sin(\gamma_++\omega_d \tau)\cos\gamma_- -Q_c^{-1} p_+ + {\cal{N}}_c (\tau)\cr
\dot{\gamma}_-&=&\frac{2 E_{C_J}}{\hbar\omega_+} (N-N_g)-\frac{4\pi E_J}{\hbar\omega_+}\left(\frac{C_g}{C_J}\right)^2 \frac{R_g}{R_K}\sin\gamma_- \cos(\gamma_+ +\omega_d \tau) +{\cal{N}}_g (\tau)\cr
\dot{N}&=&-\frac{2 E_J}{\hbar\omega_+}\sin\gamma_-\cos(\gamma_++\omega_d\tau),
\label{dimlesslangevineq}
\end{eqnarray}
where the dimensionless conversions are $\tau=\omega_+ t$ and $\tilde{p}_+=p_+/(I_+\omega_+)$, with $\omega_+=1/\sqrt{LC}$. The dimensionless drive force amplitude and frequency are $f=\pi L I_c/\Phi_0=4(Z/R_K) E_J/(\hbar\omega_+)$ and $\tilde{\omega}_d=(L/L_b) e V_{\mathrm{dc}}/(\hbar\omega_+)$, respectively [with the tildes subsequently dropped in Eq.~(\ref{dimlesslangevineq}) and below]. The cavity mode quality factor is $Q_c=R_c \sqrt{C/L}$ in terms of the cavity mode resistance $R_c$, while $R_g$ denotes the gate voltage resistance. The associated dimensionless cavity and gate bias noise ``forces" satisfy the respective correlation relations
\begin{equation}
\left\langle {\cal{N}}_c(\tau){\cal{N}}_c(0)\right\rangle=2\left(\frac{\pi}{\Phi_0}\right)^2 L k_B T_c Q_c^{-1}\delta(\tau)
\label{cavitynoisecorreq}
\end{equation}
and
\begin{equation}
\left\langle {\cal{N}}_g(\tau){\cal{N}}_g(0)\right\rangle=\frac{4\pi k_B T_g}{\hbar\omega_+} \left(\frac{C_g}{C_J}\right)^2\frac{R_g}{R_K} \delta(\tau),
\label{gatenoisecorreq}
\end{equation}
where we distinguish the cavity mode environment and gate voltage effective noise temperatures, since they are not necessarily the same in experiment.

\section{Classical dynamics}
\label{sec:classd}

The set of Langevin equations~(\ref{dimlesslangevineq}) provides a full description of the classical stochastic dynamics of the system. Numerical integration of these equations averaged over many different realizations of the noise allows one to obtain probability distributions for all of the system variables. Ultimately these distributions could then be compared with appropriately chosen quasiprobability distributions for the corresponding quantum degrees of freedom. However, this approach is rather demanding from a computational point of view, especially for the quantum dynamics. We will restrict ourselves to outlining the behavior of the simpler system consisting of the driven Cooper-pair transistor alone. In effect this corresponds to the limit of small  $f$, $T_c$ and $Q_c$.

Looking at Eq.\ (\ref{dimlesslangevineq}), it is clear that for a strongly damped and weakly driven cavity, the variable $p_+$ will remain small and hence to a good approximation it will be possible to drop the $p_+$ dependence of the $N,\gamma_-$ equations so that the latter become entirely decoupled from the evolution of the cavity variables. In this limit we are left with just the pair of equations,
\begin{eqnarray}
\dot{\gamma}_-&=&\frac{2 E_{C_J}}{\hbar\omega_+} (N-N_g)-\frac{4\pi E_J}{\hbar\omega_+}\left(\frac{C_g}{C_J}\right)^2 \frac{R_g}{R_K}\sin\gamma_- \cos(\omega_d\tau) +{\cal{N}}_g (\tau)\cr
\dot{N}&=&-\frac{2 E_J}{\hbar\omega_+}\sin\gamma_-\cos(\omega_d\tau).
\label{dimlesslangevineq2}
\end{eqnarray}

\begin{figure}
\center{
\includegraphics[width=6.5cm]{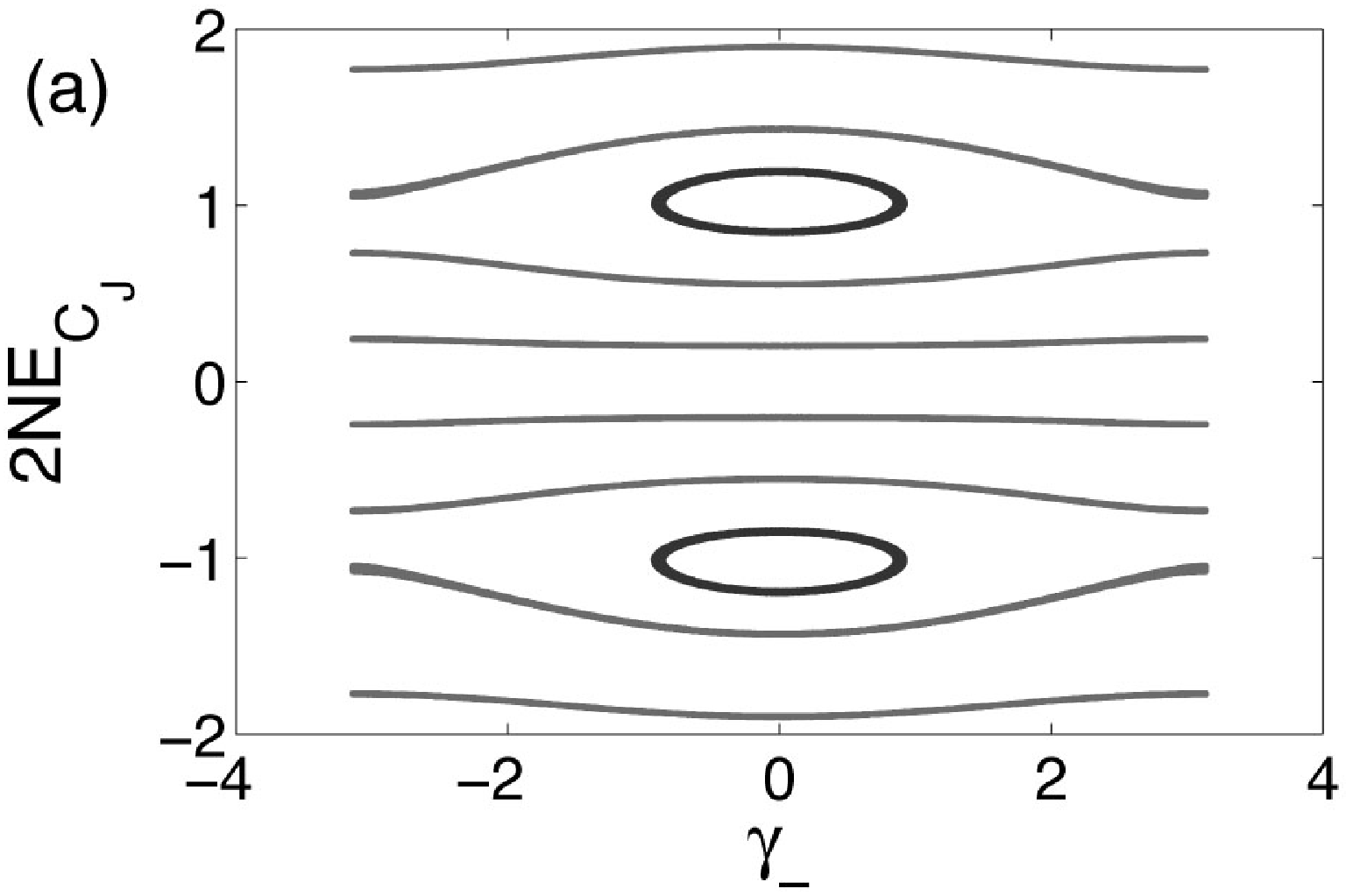}
\includegraphics[width=6.5cm]{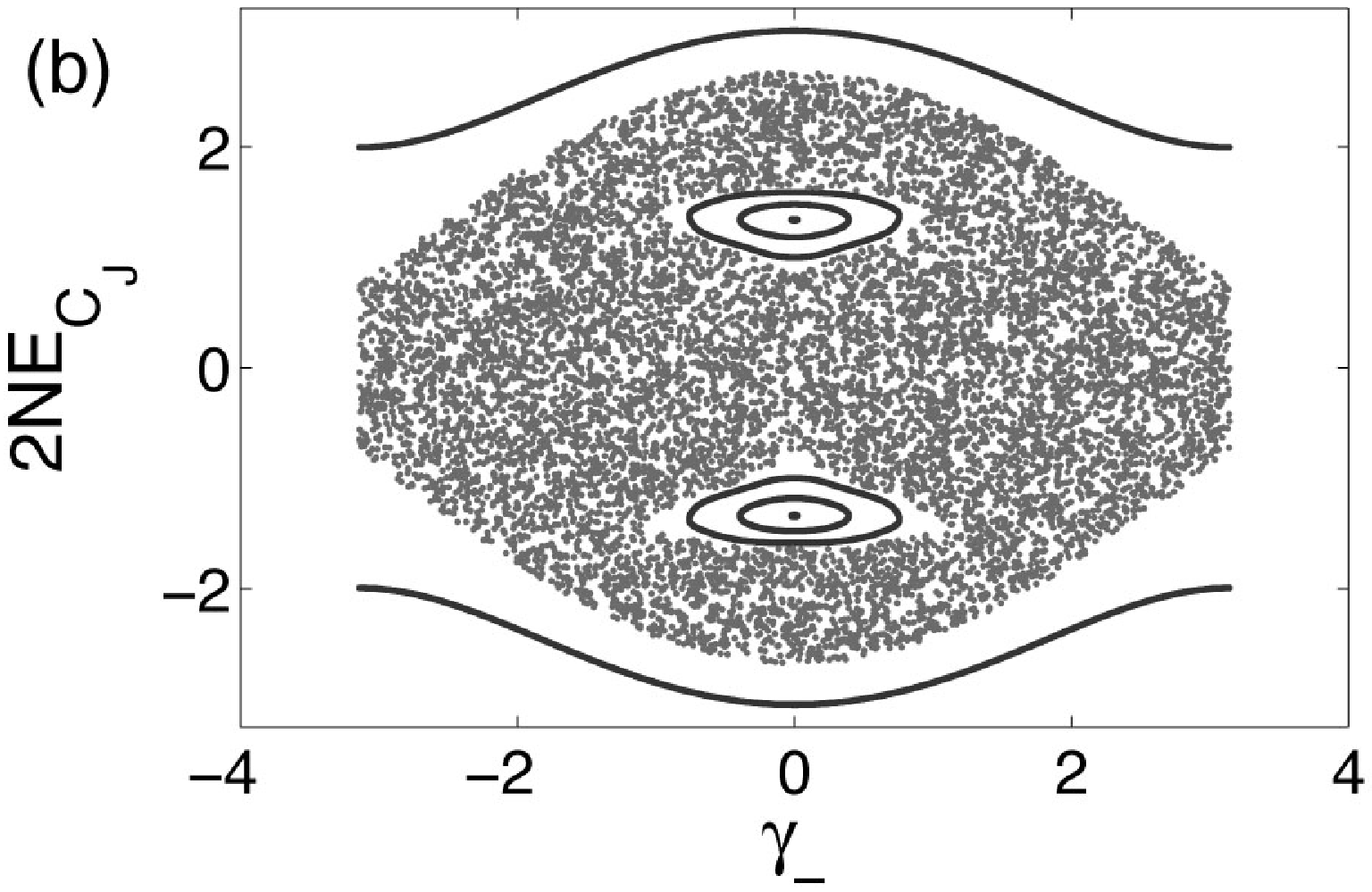}}
 \caption{Stroboscopic phase space portraits in the absence of damping. In (a) $E_J=\cos(1.55)\approx 0.021$, $E_{C_J}=\sin(1.55)\approx 1.000$ while in (b) $E_J=\cos(0.3)\approx 0.955$, $E_{C_J}=\sin(0.3)\approx 0.296$. The other parameters are $\omega_d=1$, $N_g=0$, $R_g=0$. (For numerical calculations, energies are measured in units where $\hbar\omega_d=1$)}\label{fig:class1}
\end{figure}

We start by solving Eq.\ (\ref{dimlesslangevineq2}) in the limit where $R_g=0$. In this regime the equations are simple classical equations of motion for $N$ and $\gamma_-$.  Nevertheless, they reveal a complex dynamical behavior which has already been investigated in different contexts (see e.g.\ \cite{mouchetpre06}). Depending on the initial conditions and the choice of parameters, the system typically has a mixed phase space in which the behavior is either chaotic or quasiperiodic. The phase space is visualized in a stroboscopic plot in which a point is plotted after each period of the drive, examples of which are shown in Fig.\ \ref{fig:class1}. In the limit $E_J\rightarrow 0$ the system is integrable with natural frequencies $2E_{C_J}N$, hence for very small values of $E_J/E_{C_J}$ the phase space is perturbed around resonances~\cite{reichl04} which occur at $N=\pm \omega_d/(2E_{C_J})$ (see Fig.\ \ref{fig:class1}a); as $E_J/E_{C_J}$  is increased the resonances get larger and a chaotic sea forms when they overlap. Islands of stability (where the orbits remain quasiperiodic) are found near $N=\pm \omega_d/(2E_{C_J})$ even when  $E_J/E_{C_J}>1$ (see Fig.\ \ref{fig:class1}b).

We can explore the sensitivity of the system to dissipation (as opposed to noise) by setting $T_g=0$ and changing the value of $R_g$. We find that even rather low levels of dissipation can have a significant effect on the the long time behavior. For example, for the parameters used in Fig.\ \ref{fig:class1}b with $R_g\simeq 50\Omega$, the phase space appears to contain only two attractive fixed points (one associated with each of the resonances). However, the chaotic sea is present as a transient, albeit one which can be rather long-lived: for certain initial conditions it only disappears after $>10^3$ periods of the drive.

\begin{figure}
\center{
\includegraphics[width=6.5cm]{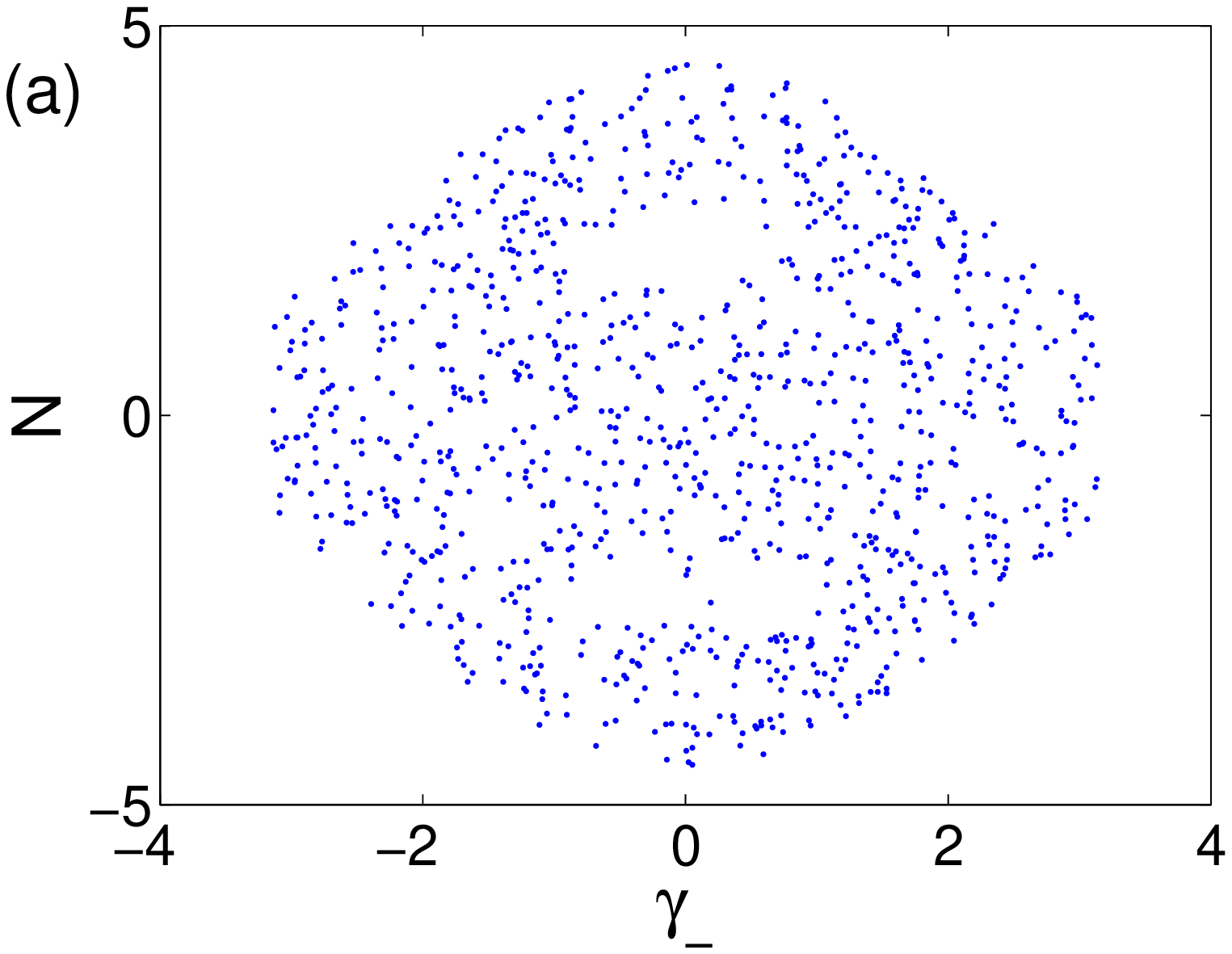}
\includegraphics[width=6.5cm]{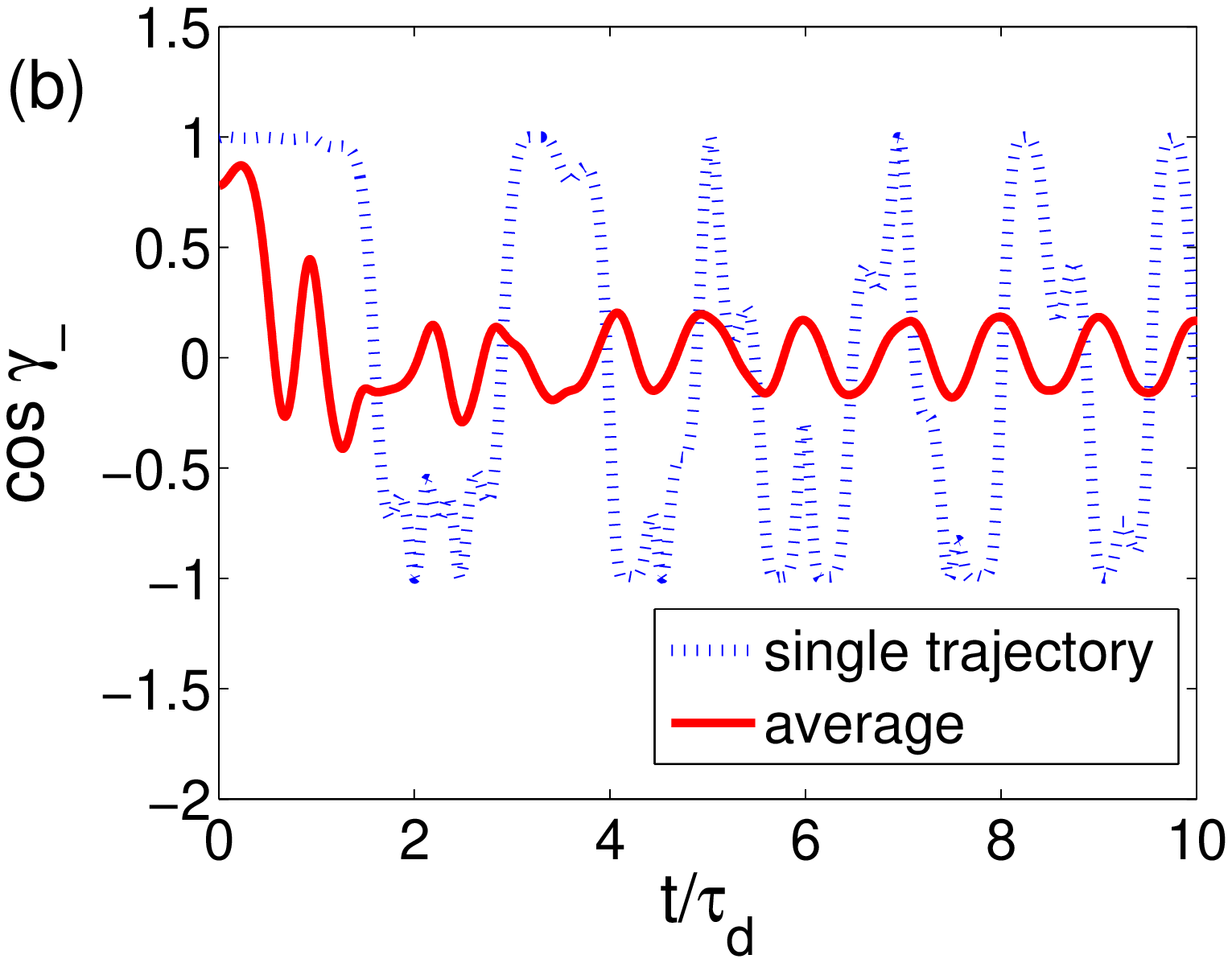}}
 \caption{Effect of a distribution of initial conditions. (a) Location in phase space of 1000 trajectories at $\tau=25\tau_d$ with initial conditions chosen from a Gaussian distribution with $\langle N\rangle=\langle\gamma_-\rangle=0$, $\Delta N=\Delta\gamma_-=1/\sqrt{2}$ (b) Evolution of $ \cos\gamma_-$ for a single trajectory starting at $\langle N\rangle=\langle\gamma_-\rangle=0.1$ compared with an ensemble of 5000 trajectories with a Gaussian distribution of initial conditions centered on the same point. The parameters are $E_J=\cos(0.3)$, $E_{C_J}=\sin(0.3)$, $\omega_d=1$, $N_g=0$, $R_g=0$.}\label{fig:class2}
\end{figure}

Before examining the full behavior of Eq.\ (\ref{dimlesslangevineq2}) with dissipation and noise, it is also worth considering the effect of averaging over an ensemble of initial conditions. In order to make a comparison with the quantum dynamics we need to consider how an initial  {\it distribution} of $N,\gamma_-$ values evolves. Because of the chaotic behavior of the system the effects of considering a range of initial coordinates can be very dramatic even after a relatively short period of time. Starting from a Gaussian distribution of initial states centered on a point in the chaotic sea, leads to a set of trajectories that spreads out rapidly over the chaotic sea as can be seen in Fig.\ \ref{fig:class2}a. The islands within the chaotic sea stand out (the handful of points that lie within the islands come from initial points that didn't fall within the chaotic sea). Clearly averaging over a range of initial conditions has a dramatic effect on the dynamics of the averages of the system, this is particularly clear for the quantity $\langle \cos\gamma_-\rangle$ which very rapidly becomes a periodic oscillation with period $\tau_d=2\pi/\omega_d$ as shown in Fig.\ \ref{fig:class2}b.

\begin{figure}
\center{
\includegraphics[width=6.5cm]{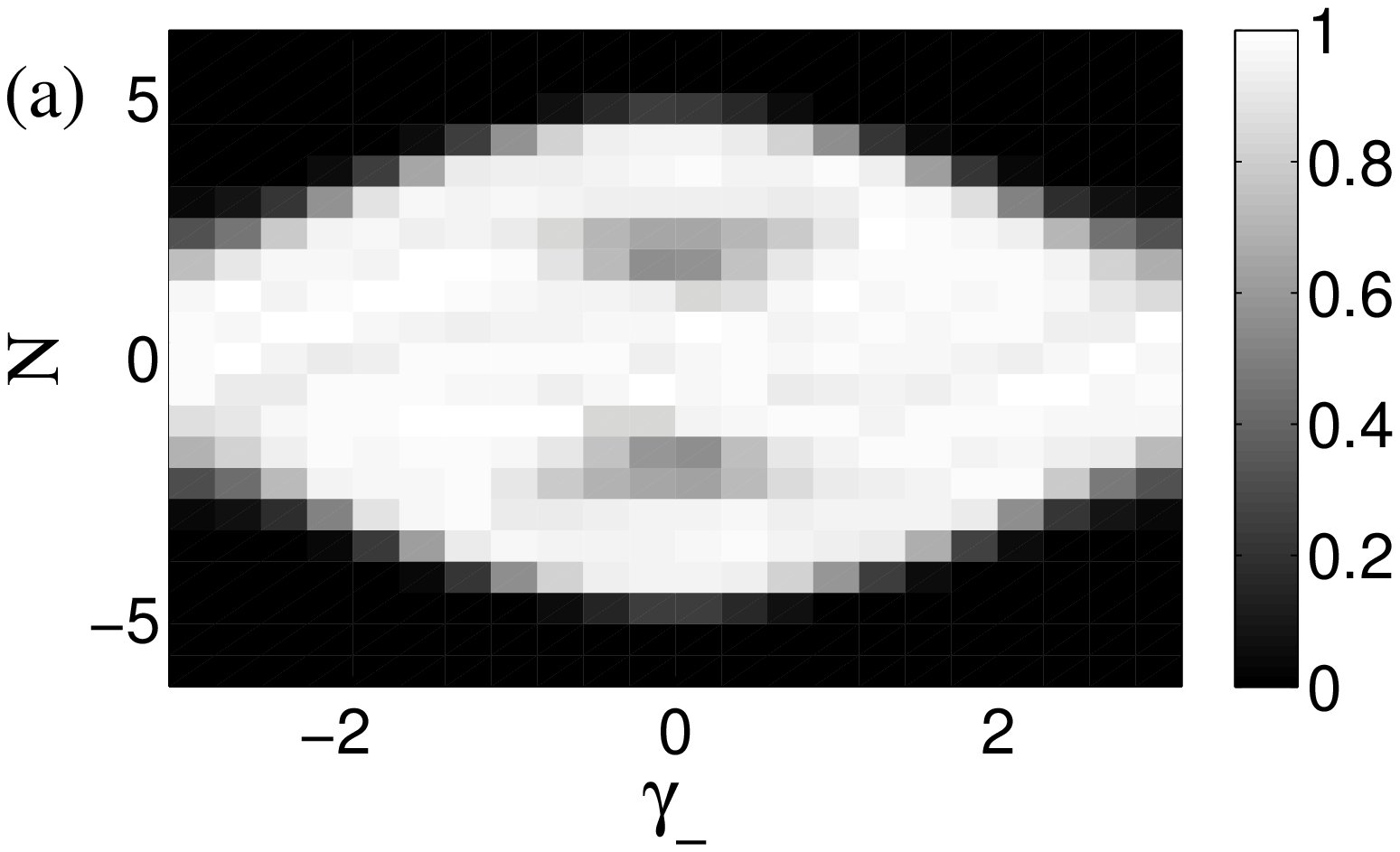}
\includegraphics[width=6.5cm]{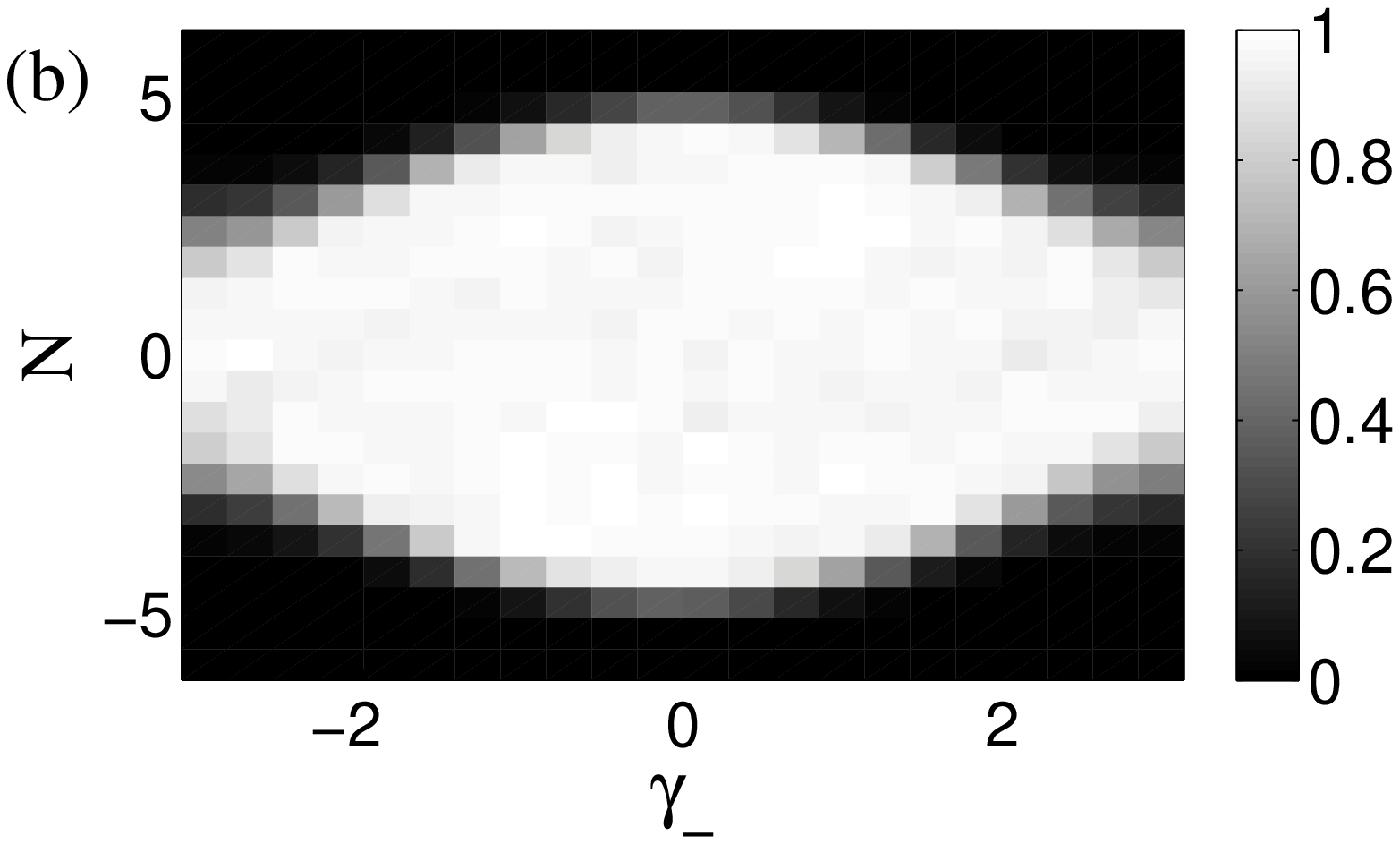}}
 \caption{Stochastic probability distributions at times (a) $\tau=10\tau_d$ and (b) $\tau=25\tau_d$ (the scale gives the probability of each square relative to the maximum). The parameters are $E_J=\cos(0.3)$, $E_{C_J}=\sin(0.3)$, $\omega_d=1$, $N_g=0$, $R_g=500\Omega$ and $k_{\rm B}T_g=2E_J$. Note we take $C_g/C_J=0.1$ throughout.}\label{fig:stoch1}
\end{figure}

Examples of the probability distribution for the classical, noisy, evolution of the system are shown in Fig.\ \ref{fig:stoch1}. The numerical interaction is carried out using a generalization of the Heun method used for deterministic differential equations \cite{breuer}.  In this case an average is carried out both over realizations of the noise and the initial conditions which are chosen from a Gaussian distribution with variances $\Delta N=\Delta\gamma_-=1/\sqrt{2}$ centered on a given point in phase space. When noise is added to the system the trajectories eventually diffuse between the chaotic sea and the quasiperiodic orbits so that the difference in the probability distribution over the island and chaotic sea regions gets washed out over time. In Fig.\ \ref{fig:stoch1} the remnants of the island can be seen at $\tau=10\tau_d$, but by $\tau=25\tau_d$ they have disappeared completely.

\section{Quantum model of device}
\label{sec:quantumeq}

\subsection{Quantum master equation}
The Poisson bracket relations for the classical canonical coordinates are
\begin{equation}
\left\{\gamma_+,p_+\right\}=1;~\left\{\gamma_-,N\right\}=\hbar^{-1}
\label{poissoneq}
\end{equation}
(where recall $N=p_-/\hbar$).
Applying the correspondence principle, the quantum commutation relations are
\begin{equation}
[\hat{\gamma}_+,\hat{p}_+]=i\hbar;~[\hat{\gamma}_-,\hat{N}]=i.
\label{commutatoreq}
\end{equation}
However,  the phase coordinates are not periodic functions of their associated system configuration spaces;  the representations of the commutation relations (\ref{commutatoreq}) give the unbounded eigenvalue spectrum  $\mathbb{R}$  for the corresponding phase operators. While this is not a problem for the `torsional' oscillator  because of the strong harmonic confining potential, which limits the accessible region of configuration space, the `pendulum'  typically explores the whole of its unit circle ($S^1$) configuration space. A suitable pendulum configuration space function is $e^{i\gamma_-}$ with Poisson bracket relation:
\begin{equation}
\left\{e^{i\gamma_-},N\right\}=i\hbar^{-1}e^{i\gamma_-}.
\label{poisson2eq}
\end{equation}
The corresponding commutation relation is then
\begin{equation}
[e^{i\hat{\gamma}_-},\hat{N}]=-e^{i\hat{\gamma}_-}.
\label{commutator2eq}
\end{equation}
Eq.~(\ref{commutator2eq}) has infinitely many unitarily inequivalent representations~\cite{kastruppra06} that can be labelled by a real parameter $0\leq\delta<1$. Each representation is spanned by a number basis $|N\rangle_{\delta}$, where
\begin{equation}
\hat{N} |N\rangle_{\delta}=(N+\delta)|N\rangle_{\delta}, N=0,\pm1,\pm2,\dots.
\label{deltarepeq}
\end{equation}
Introduce raising and lowering operators for the torsional (CoM) coordinate:
\begin{equation}
\gamma_+=\sqrt{\frac{\hbar}{2I_+\omega_+}}(a+a^+);~p_+=i\sqrt{\frac{I_+\omega_+\hbar}{2}}(a^+-a),
\label{raiselowereq}
\end{equation}
where recall  $\omega_+=(LC)^{-1/2}$,  the `moment of inertia' is $I_+=C(\Phi_0/\pi)^2$, and we have dropped the hats on the operators for notational convenience. The CoM phase coordinate oscillator zero-point uncertainty is
\begin{equation}
\Delta^+_{zp}=\sqrt{\frac{\hbar}{2I_+\omega_+}}=\sqrt{\frac{\pi\sqrt{L/C}}{R_K}}=\sqrt{\frac{Z}{R_K}},
\label{zpeq}
\end{equation}
where recall $Z\approx 50~\Omega$ is the cavity impedance and $R_K=h/e^2 \approx 25.8~{\mathrm{k}}\Omega$ is the von Klitzing constant, so that $\Delta^+_{zp}=(50/25800)^{1/2}\approx 0.04$.
The Hamiltonian operator corresponding to (\ref{transformedhamiltonian3eq}) is
\begin{eqnarray}
&&{\cal H}=\hbar\omega_+ a^+a +E_{C_J}\sum_{N=-\infty}^{+\infty}\left(N+\delta-N_g\right)^2|N\rangle_{\delta \delta}\langle N|\cr
&&-E_J\sum_{N=-\infty}^{+\infty}\left(|N+1\rangle_{\delta \delta}\langle N|+|N-1\rangle_{\delta \delta}\langle N|\right)\cos\left[\Delta^+_{zp} (a+a^+) +\omega_d t\right].\cr
&&\label{quantumhamiltonianeq}
\end{eqnarray}
The parameter $\delta$ appearing in the Hamiltonian operator is a purely quantum signature of the nontrivial topology of the corresponding classical pendulum's  configuration space $S^1$. An interesting question concerns the particular value for $\delta$ that Nature chooses and why~\cite{kastruppra06,kowalski,bahr}. However, it is likely not possible to measure $\delta$ in experiment, since from (\ref{quantumhamiltonianeq}) it is clear that the effect of a nonzero $\delta$ value is indistinguishable from that due to the presence of an excess charge on the CPT island. From now on, we shall set $\delta=0$.

We now derive the open system quantum master equations within the self-consistent Born approximation (SCBA) following the approach reviewed in Ref.~\cite{paz01}. In the following, we analyze the two noise/dissipation sources independently, beginning first with the cavity mode environment.  We  write the Hamiltonian~(\ref{transformeddisshamiltonian2eq}) as ${\cal{H}}={\cal{H}}_S +{\cal{H}}_E +{\cal{V}}$ where the Hamiltonian ${\cal{H}}_S$ describes the cavity mode-CPT system [Eq.~(\ref{quantumhamiltonianeq})], the Hamiltonian ${\cal{H}}_E$ describes the  infinite parallel $LC$ network environment, and the interaction part is
\begin{equation}
{\cal{V}}=-2\left(\frac{\Phi_0}{2\pi}\right)^2\gamma_+ \sum_i \frac{1}{L_i}\phi_i=-\gamma_+\sum_i\lambda_i\phi_i,
\label{interactionparteq}
\end{equation}
where $\lambda_i=2(\Phi_0/(2\pi))^2(L_i)^{-1}$.
Defining $B=\sum_i\lambda_i\phi_i$, we obtain for the master equation within the SCBA:
\begin{eqnarray}
\dot{\rho}(t)&=&-\frac{i}{\hbar}\left[{\cal{H}}_S ,\rho(t)\right] \cr
&&-\frac{1}{\hbar^2}\int_0^t dt' \left\{\frac{1}{2}\langle\left\{B(t),B(t')\right\}\rangle\left[\gamma_+,\left[\gamma_+(t'-t),\rho(t)\right]\right]\right.\cr
&&+\left.\frac{1}{2}\langle\left[B(t),B(t')\right]\rangle\left[\gamma_+,\left\{\gamma_+(t'-t),\rho(t)\right\}\right]\right\},
\label{detailedcavitymastereq}
\end{eqnarray}
where the operators $B(t)$ and $\gamma_+ (t'-t)$ are in the interaction picture and the expectation values are performed assuming the environment (infinite parallel $LC$ network) is in a thermal state. We have
\begin{eqnarray}
\frac{1}{2}\langle\left\{B(t),B(0)\right\}\rangle&=&\frac{1}{\pi}\int_0^{\infty}d\omega J(\omega) \cos\omega t \left(1+2 N(\omega)\right)\cr
\frac{1}{2}\langle\left[B(t),B(0)\right]\rangle&=&-\frac{i}{\pi} \int_0^{\infty}d\omega J(\omega) \sin\omega t
\label{bathcorrelationseq},
\end{eqnarray}
where $N(\omega)=[\exp(\hbar\omega/k_B T_c)-1]^{-1}$ is the thermal occupation number of the environment at frequency $\omega$ and where the spectral function is
\begin{equation}
J(\omega)=\sum_i  \frac{\pi\hbar}{2 m_i\omega_i}\lambda_i^2\delta (\omega-\omega_i),
\label{spectralfunctioneq}
\end{equation}
with $m_i=C_i (\Phi_0/(2\pi))^2$ and $\omega_i=1/\sqrt{L_iC_i}$.

In Sec.~\ref{classicallimitsec}, we compare the quantum versus classical dynamics and establish conditions under which the former is well approximated by the latter--the so-called classical limit. A necessary condition to be in the classical limit is that the environment temperature must be sufficiently large such that we can make the approximation $N(\omega)\approx k_B T_c/(\hbar\omega)\gg 1$. This requires $k_B T_c\gg\hbar\omega_+$. The environment correlation function then becomes
\begin{equation}
\frac{1}{2}\langle\left\{B(t),B(0)\right\}\rangle=k_B T_c \sum_i \frac{\lambda_i^2}{m_i\omega_i^2}\cos\omega_i t = \left(\frac{\Phi_0}{\pi}\right)^2 C k_B T_c\ \Gamma(t),
\label{highTcorreq}
\end{equation}
where $\Gamma(t)$ is the classical damping kernel (\ref{kerneleq}). If, furthermore, the spectral function upper cut-off satisfies $\Lambda\gg k_B T_c$, then we can make the Markovian approximation $\Gamma (t) \approx \frac{2}{RC}\delta(t)$ and
\begin{equation}
\frac{1}{2}\langle\left[B(t),B(t')\right]\rangle=-i \frac{\hbar}{2}\left(\frac{\Phi_0}{\pi}\right)^2 C \frac{d}{dt'} \Gamma(t-t')\approx -i \frac{\hbar}{2}\left(\frac{\Phi_0}{\pi}\right)^2 C \cdot\frac{2}{RC}\frac{d}{dt'}\delta(t-t').
\label{dampcorreq}
\end{equation}
Substituting expressions~(\ref{highTcorreq}) and (\ref{dampcorreq})  into Eq.~(\ref{detailedcavitymastereq}), integrating by parts and using $p_+=\left(\Phi_0/\pi\right)^2 C\dot{\gamma}_+$, we obtain
\begin{equation}
\dot{\rho}(t)=-\frac{i}{\hbar} \left[ {\cal{H}}_S,\rho(t)\right]-\frac{i}{2\hbar} \Gamma\left[\gamma_+,\left\{p_+,\rho(t)\right\}\right]-\frac{1}{\hbar^2}\Gamma I_+ k_B T_c \left[\gamma_+
\left[\gamma_+,\rho(t)\right]\right],
\label{standardmastereq}
\end{equation}
where $\Gamma=1/(RC)$ is the damping rate. Eq.~(\ref{standardmastereq}) is just the standard Born-Markov master equation for a quantum Brownian particle in the high temperature limit~\cite{paz01}, where the second term on the right hand side describes damping and the third term on the right hand side describes diffusion.

While Eq.~(\ref{standardmastereq}) is appropriate for investigating the classical limit, under the cryogenic conditions of an actual experiment and for say an $\omega_+\gtrsim 2\pi\times 5~{\mathrm{GHz}}$ cavity mode, we expect that  $k_B T_c\ll\hbar\omega_+$, so that a low temperature limit is more appropriate.  Using Eq.~(\ref{raiselowereq}) to express the master equation in  terms of raising and lowering operators, making the rotating wave approximation and the replacement $k_BT_c\rightarrow \hbar\omega_+/2$, we obtain the following `low temperature' master equation:
\begin{equation}
\dot{\rho}=-\frac{i}{\hbar}[{\cal{H}}_S,\rho]-\frac{1}{2}\Gamma \left(a^+a\rho +\rho a^+ a-2a\rho a^+\right).
\label{mastereq}
\end{equation}
The non-Hermitian part of the master equation~(\ref{mastereq}) is of the Lindblad form, ensuring that the solution to~(\ref{mastereq})  for the density matrix  $\rho(t)$ is consistent. However, in the above outlined derivation of this master equation, there is no apparent justification for the replacement $k_B T_c\rightarrow \hbar\omega_+/2$, since $\omega_+$ is not in general the characteristic frequency for dynamics described by the nonlinear, time-dependent Hamiltonian ${\cal{H}}_S$. Only for a harmonic oscillator Hamiltonian with classical frequency $\omega_+$ is this replacement justified, provided the damping is sufficiently weak, i.e., $\hbar \Gamma\ll k_B T_c \ll\hbar\omega_+$. Nevertheless, for better or worse, we shall follow common practise and assume that master equation~(\ref{mastereq}) provides an adequate model for weakly damped cavity mode-CPT quantum dynamics at  low temperatures.

We move on now to derive the master equation modelling gate voltage noise, starting from the Hamiltonian~(\ref{transformedgatedisshamiltonian2eq}). For the master equation within the SCBA, we obtain [c.f., Eq.~(\ref{detailedcavitymastereq})]
\begin{eqnarray}
\dot{\rho}(t)&=&-\frac{i}{\hbar}\left[{\cal{H}}_S ,\rho(t)\right] \cr
&&-\frac{1}{\hbar^2}\int_0^t dt' \left\{\frac{1}{2}\langle\left\{B(t),B(t')\right\}\rangle\left[N,\left[N(t'-t),\rho(t)\right]\right]\right.\cr
&&+\left.\frac{1}{2}\langle\left[B(t),B(t')\right]\rangle\left[N,\left\{N(t'-t),\rho(t)\right\}\right]\right\},
\label{detailedgatemastereq}
\end{eqnarray}
where now $B=\sum_i\lambda_i p_i$, with $\lambda_i =(2 E_{C_J}/\hbar)(C_g/C_i)$. The correlation functions still take the same form as Eq.~(\ref{bathcorrelationseq}), but where now the  spectral function is
\begin{equation}
J (\omega)=\sum_i \frac{\pi m_i\omega_i\hbar}{2}\lambda_i^2\delta(\omega-\omega_i).
\label{gatespectralfunction}
\end{equation}
In the high temperature limit, appropriate for comparing the quantum versus classical dynamics, the environment correlation functions become
\begin{eqnarray}
\frac{1}{2}\langle\left\{B(t),B(0)\right\}\rangle&=&k_B T_g\left(\frac{e C_g}{C_J}\right)^2 \sum_i \frac{1}{C_i}\cos\omega_i t=2 k_B T\left(\frac{e C_g}{C_J}\right)^2 R_g \delta(t)\cr
\frac{1}{2}\langle\left[B(t),B(0)\right]\rangle&=&\frac{i\hbar}{2 k_B T} \frac{d}{dt}\left[\frac{1}{2}\langle\left\{B(t),B(0)\right\}\rangle\right]=i\hbar\left(\frac{e C_g}{C_J}\right)^2 R_g\frac{d}{dt}\delta(t),\cr
&&\label{quantumgatecorreq}
\end{eqnarray}
where the last equality results from making the Markovian approximation. Substituting the correlation relations~(\ref{quantumgatecorreq}) into the master equation~(\ref{detailedgatemastereq}), integrating the damping term by parts and using also Heisenberg's equation to solve for $\dot{N}$, we obtain:
\begin{eqnarray}
\dot{\rho}(t)&=&-\frac{i}{\hbar}\left[{\cal{H}}_S ,\rho(t)\right] -\frac{1}{\hbar^2}\left(\frac{e C_g}{C_J}\right)^2 k_B T_g R_g \left[N,\left[N,\rho(t)\right]\right]\cr
&&-\frac{i}{\hbar^2}\left(\frac{e C_g}{C_J}\right)^2 E_J R_g \left[N,\left\{\sin\gamma_-\cos(\gamma_+ +\omega_d t),\rho(t)\right\}\right],
\label{markovgatemastereq}
\end{eqnarray}
where the second term on the right hand side describes diffusion and the third term describes damping. Note the atypical, explicit time-dependence in the  damping term.

In contrast with the more familiar quantum Brownian master equation~(\ref{standardmastereq}) for the cavity mode environment,  there is no corresponding simple prescription for recovering from Eq.~(\ref{markovgatemastereq}) a consistent, low temperature master equation that is appropriate for lower gate voltage noise levels expected in an actual cryogenic experiment, where the bias lines are filtered. Given the difficulties in finding such a low temperature master equation, we shall instead in Sec.~\ref{quantumdynsec} `take the path of least resistance' and simply compare the quantum dynamics in both the presence and absence of the environment non-Hermitian terms in (\ref{markovgatemastereq}), so as to gain some understanding of the system quantum dynamics in the presence of gate voltage noise.

\subsection{Wigner functions and coherent states}
\label{wignersec}
In order to make a comparison between the quantum and classical dynamics of our systems we need to identify suitable tools with which to describe the quantum dynamics. For continuous systems coherent states prove useful as initial states because they are localized in phase space and have minimum uncertainty making them the quantum states most closely connected to a classical phase space point. Furthermore, the Wigner function quasiprobability distribution provides an effective way of visualizing the quantum evolution of a continuous variable system in phase space and signals the presence of quantum interference effects by turning negative. Whilst both conventional coherent states and the Wigner function can be applied directly to the cavity degrees of freedom, the Cooper-pair transistor  is different since the Cooper pair number is discrete. Nevertheless one can define appropriate versions of both coherent states and Wigner functions for the $N,\gamma_-$ degrees of freedom, but with some important differences compared to the usual continuous variable case.

Given the basic algebra of our number and phase operators, $[\hat{\gamma}_-,\hat{N}]=i$ and $[\hat{N},{\rm e}^{i\hat{\gamma}_-}]={\rm e}^{i{\gamma}_-}$, we can use a definition of coherent states first developed for angular momentum and rotation angle variables. We define the coherent states \cite{kowalski,bahr} as eigenstates of the operator $\hat{X}={\rm e}^{i\hat{\gamma}_--\hat{N}}$ and hence they take the form
\[
|\chi\rangle=\frac{1}{n^{1/2}}\sum_{j=-\infty}^{+\infty}\chi^{-j} {\rm e}^{-j^2/2} |j\rangle,
\]
where the normalization factor is
\[
n=\sum_{j=-\infty}^{+\infty} |\chi|^{-2j}{\rm e}^{-j^2}.
\]
The complex parameter $\chi$ can be written in terms of a charge ${N}$ and phase ${\gamma}_-$, $\chi={\rm e}^{i{\gamma}_--{N}}$, which are closely related to the expectation values of the corresponding operators \cite{kowalski},
\begin{equation}
\langle \chi |\hat{N}|\chi\rangle\simeq {N}
\end{equation}
and
\begin{equation}
\langle \chi |{\rm e}^{i\hat{\gamma}_-}|\chi\rangle\simeq {\rm e}^{-1/4}{\rm e}^{i{\gamma}_-}.
\end{equation}

Although the properties of these states differ somewhat from those of the harmonic oscillator coherent states they nevertheless take a fairly simple and intuitive form in phase space. Again, following the approach developed for angular momentum and rotation angle variables, we can use the form of the Wigner function developed for this case~\cite{berry,mukunda,bizarro}. For a system with density operator ${\rho}$ the quasiprobability distribution is given by~\cite{bizarro},
\begin{eqnarray}
W_N(\gamma_-,t)&=&\frac{1}{\pi}\int_{-\pi/2}^{+\pi/2} d\gamma'_- e^{-2iN\gamma'_-}\langle\gamma_- +\gamma'_- |\rho|\gamma_- -\gamma'_-\rangle\cr
&=&\frac{1}{2}\sum_{\mu=0,1}\sum_{N'=-\infty}^{+\infty} \frac{\sin\left[\left(N-N'-\mu/2\right)\pi\right]}{\left(N-N'-\mu/2\right)\pi} w_{N'+\mu/2} (\gamma_-,t),
\label{circlewignereq}
\end{eqnarray}
where
\begin{eqnarray}
 w_{N+\mu/2} (\gamma_-,t)&=&\frac{1}{\pi}\sum_{N'=-\infty}^{+\infty} e^{-2i(N'+\mu_-/2) \gamma_-} \langle N-N'|\rho|N+N'+\mu\rangle\cr
 &=&\frac{1}{\pi}\int_{-\pi}^{+\pi}d\gamma'_- e^{-2i(N+\mu_-/2) \gamma'_-} \langle \gamma_- -\gamma'_-|\rho|\gamma_- +\gamma'_-\rangle
 \label{parityfunctioneq}
 \end{eqnarray}
and $|\gamma_-+\gamma_-'\rangle$ is one of the phase states. Note that the $w$ functions have definite parity, i.e., $w_{N+\mu/2}(\gamma_- +\pi,t)=(-1)^{\mu} w_{N+\mu/2}(\gamma_- ,t)$.
Thus the Wigner function takes the form of a discrete series of strips labeled by the Cooper-pair number and which are continuous in the phase.

\begin{figure}
\center{
\includegraphics[width=6.5cm]{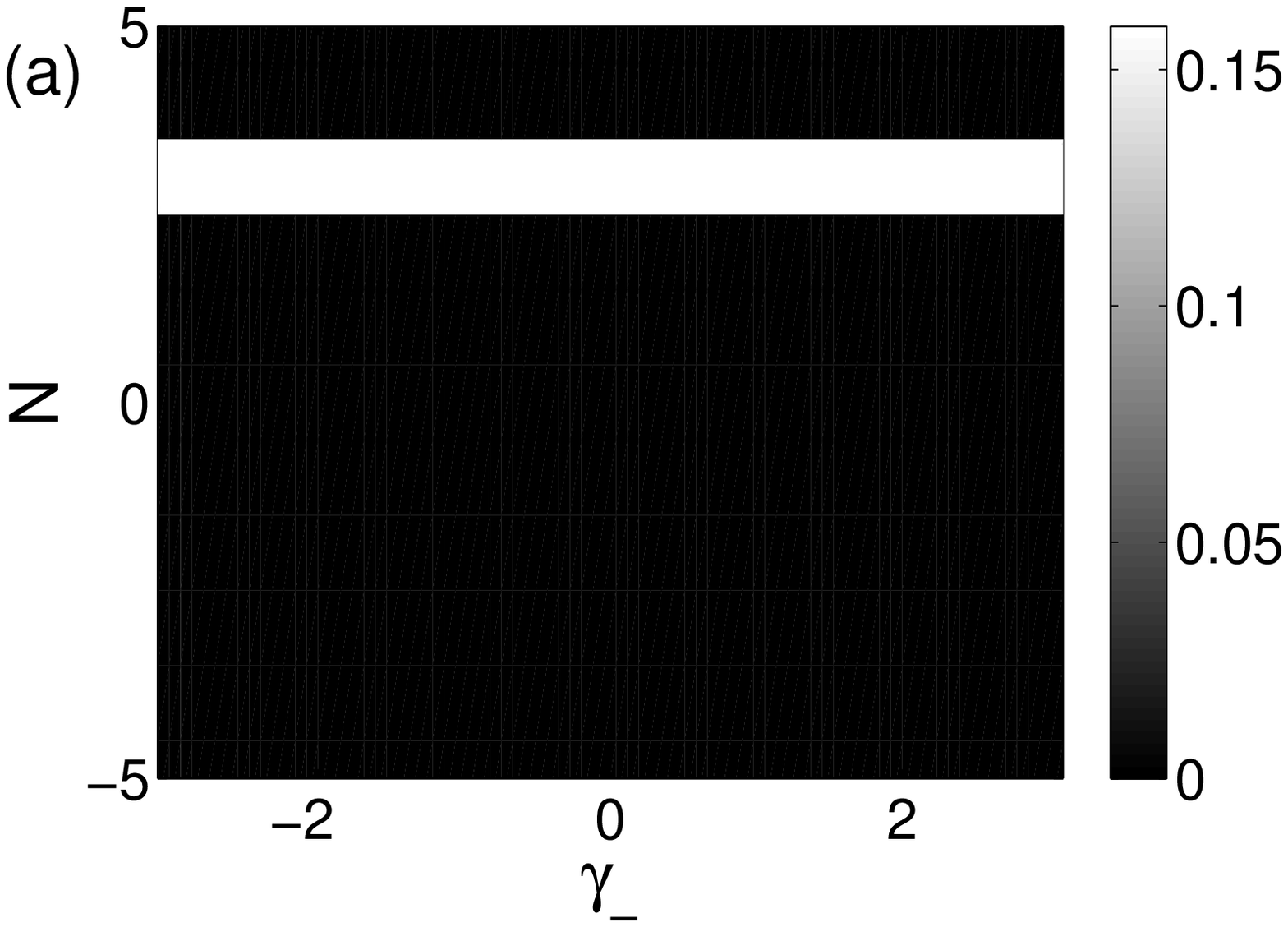}
\includegraphics[width=6.5cm]{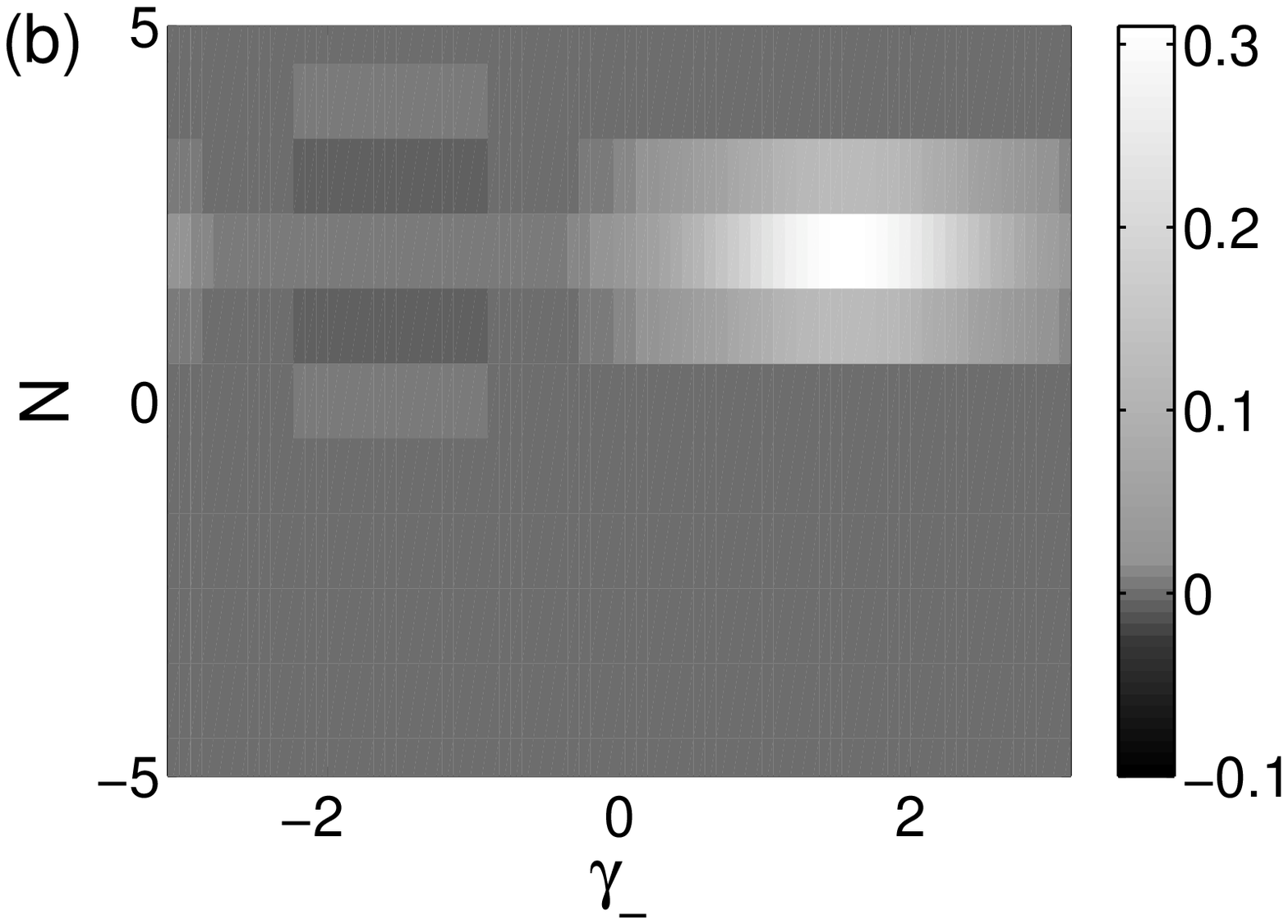}}
 \caption{Wigner functions for  a charge state $|N=3\rangle$ (a) and a coherent states $|\chi={\rm e}^{i\pi/2-2}\rangle$ (b). Note that in (b) there are regions around $\gamma_-=-\pi/2$ and $N=1,3$ where the Wigner function becomes very slightly negative.}\label{fig:exwfns}
\end{figure}

Examples of Wigner functions for a charge state $|N=3\rangle$ and the coherent state $|\chi={\rm e}^{i\pi/2-2}\rangle$ are shown in Fig.\ \ref{fig:exwfns}. The Wigner function for a charge state, $|N=M\rangle$, takes a very simple form: $W_N(\gamma_-)=\delta_{N,M}/(2\pi)$, i.e., it is just a uniform strip for $N=M$ and zero elsewhere. Interestingly the charge states are the only examples of pure states which give rise to Wigner functions which are positive everywhere \cite{rigas}. For number-phase variables the Wigner functions of coherent states do have negative regions, in contrast with the situation for continuous variable systems \cite{hudson}. However, if we wish to consider an initial quantum state which is analogous to a classical point in phase space the coherent states are still a very good choice as the amount of negativity which actually occurs is in fact very small in practice [See Fig.\ \ref{fig:exwfns}b] and our only choice if we wished to eliminate the negativity entirely while still using a pure state is to work with charge states which are completely spread out in phase. As we can see from Fig.\ \ref{fig:exwfns}b the Wigner function for $|\chi={\rm e}^{i\pi/2-2}\rangle$ is strongly peaked around $N=2,\gamma_-=\pi/2$ and, apart from very small regions of negativity \cite{rigas2008}, is very reminiscent of the corresponding continuous variable case.

\section{Quantum dynamics}
\label{quantumdynsec}

In exploring the quantum dynamics we again choose to focus on just the behavior of the Cooper-pair transistor, as we did with the classical dynamics. This in effect means that we take the limit $\Delta^+_{zp}\rightarrow 0$ in the Hamiltonian [Eq.\ (\ref{quantumhamiltonianeq})]. The presence in the model of a non-zero gate resistance and large effective gate voltage noise temperature ensures that the long-time behavior of the CPT system is classical for sufficiently large $E_J$, in the sense that the Wigner function will be everywhere positive and also a smoothly varying function of $N$ (see Sec.~\ref{classicallimitsec}). However, over short times a very different picture emerges: even very `classical' choices of the initial state can evolve naturally into states with strongly non-classical features.

\begin{figure}
\center{
\includegraphics[width=6.5cm]{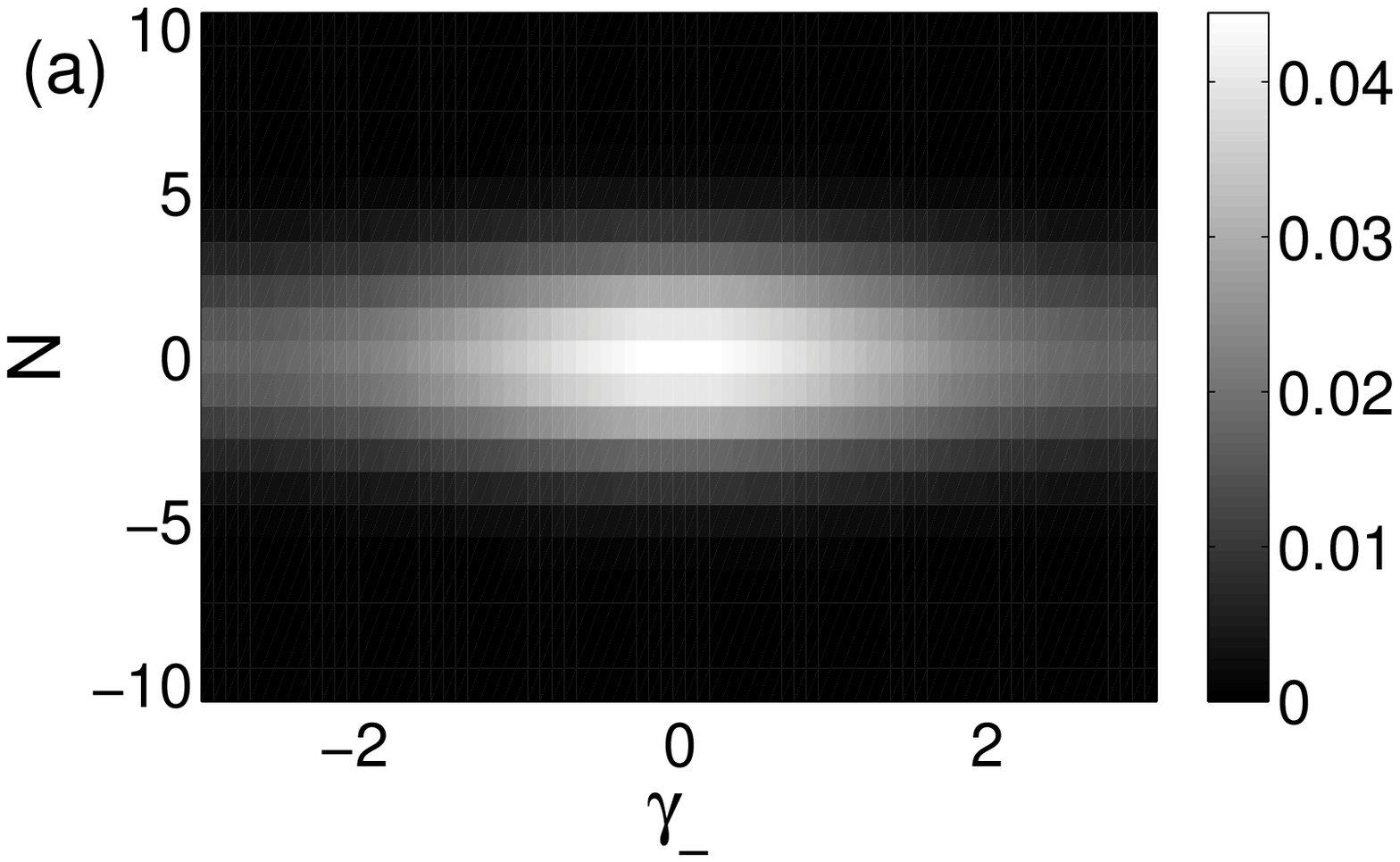}
\includegraphics[width=6.5cm]{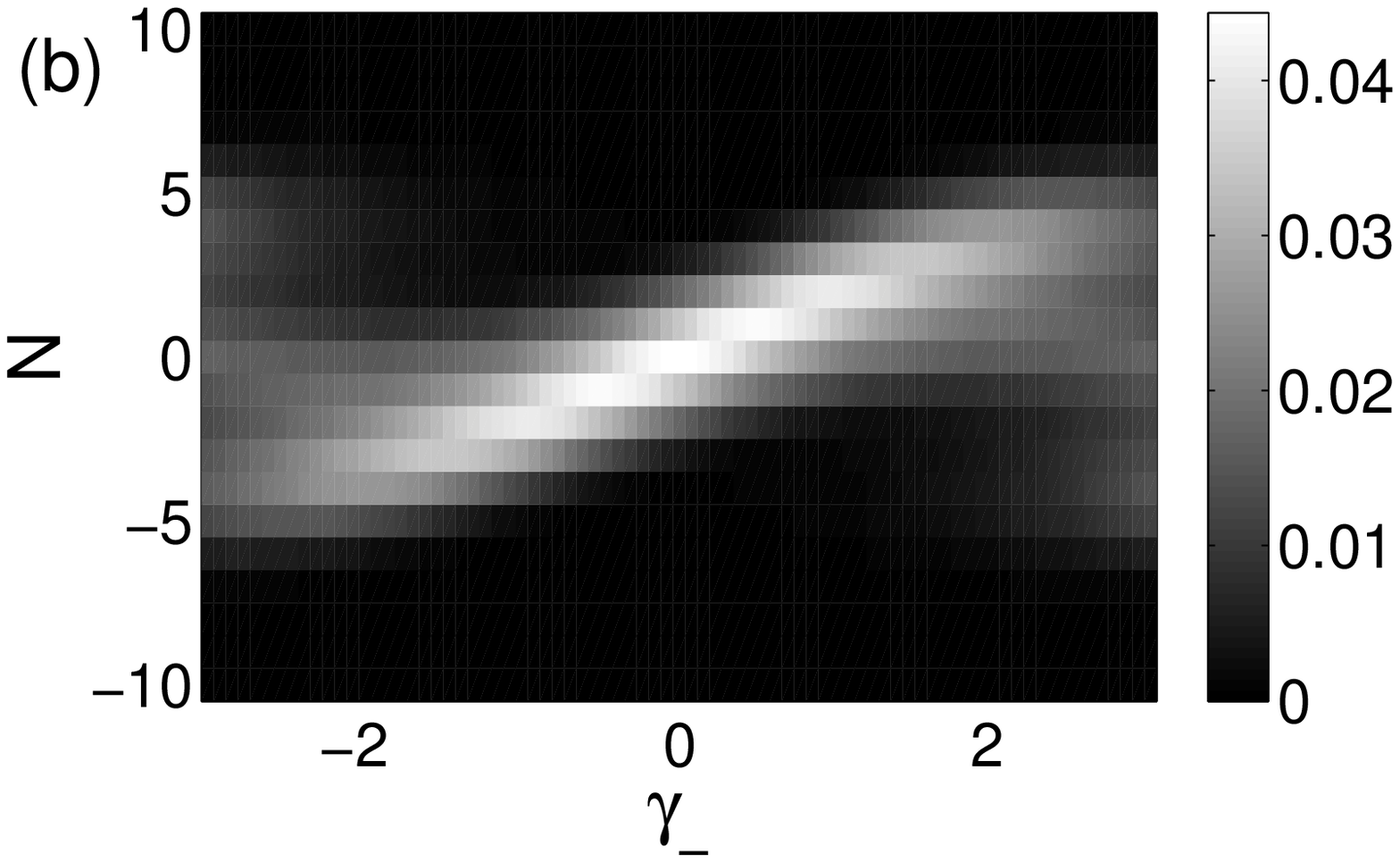}
\includegraphics[width=6.5cm]{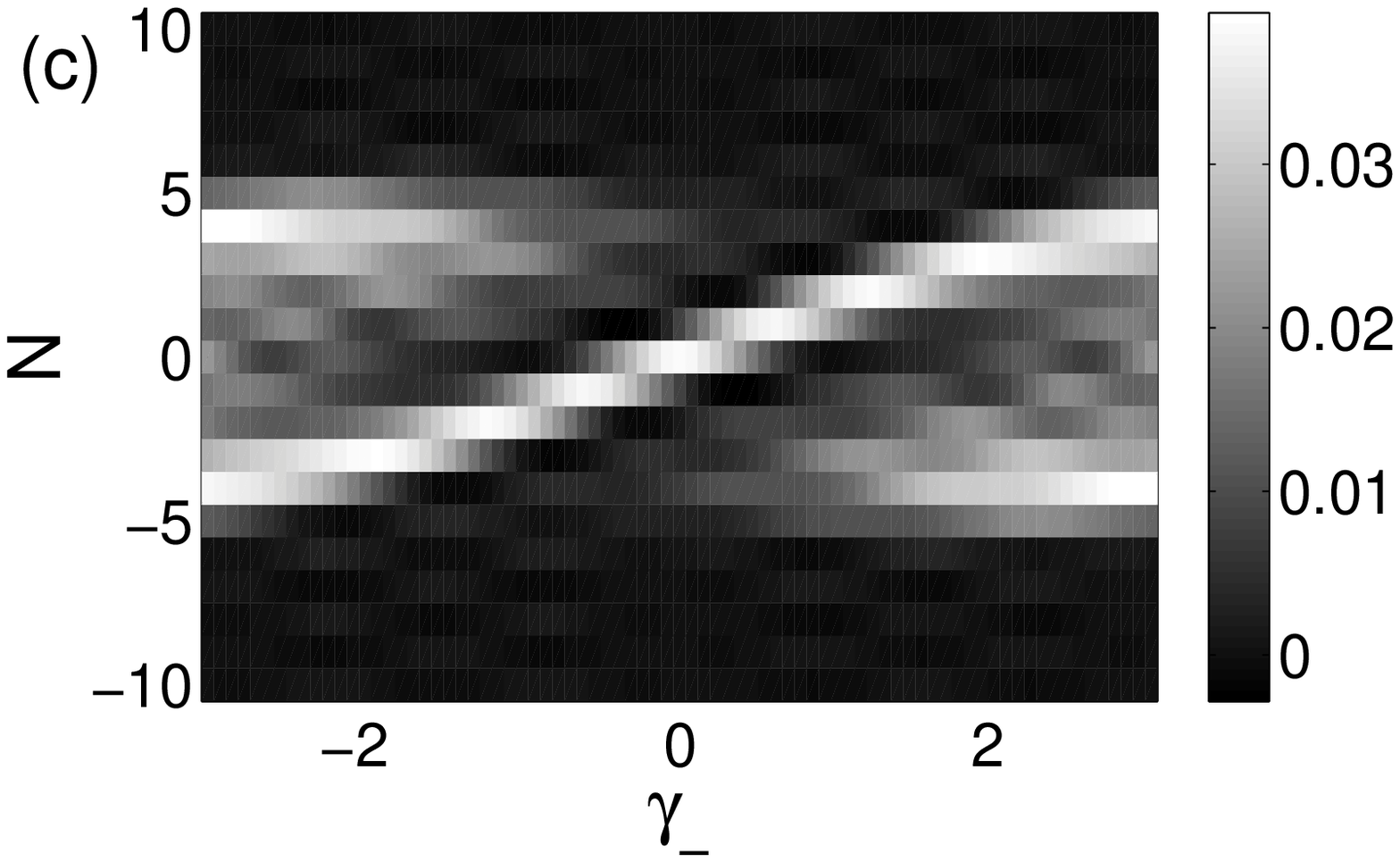}
\includegraphics[width=6.5cm]{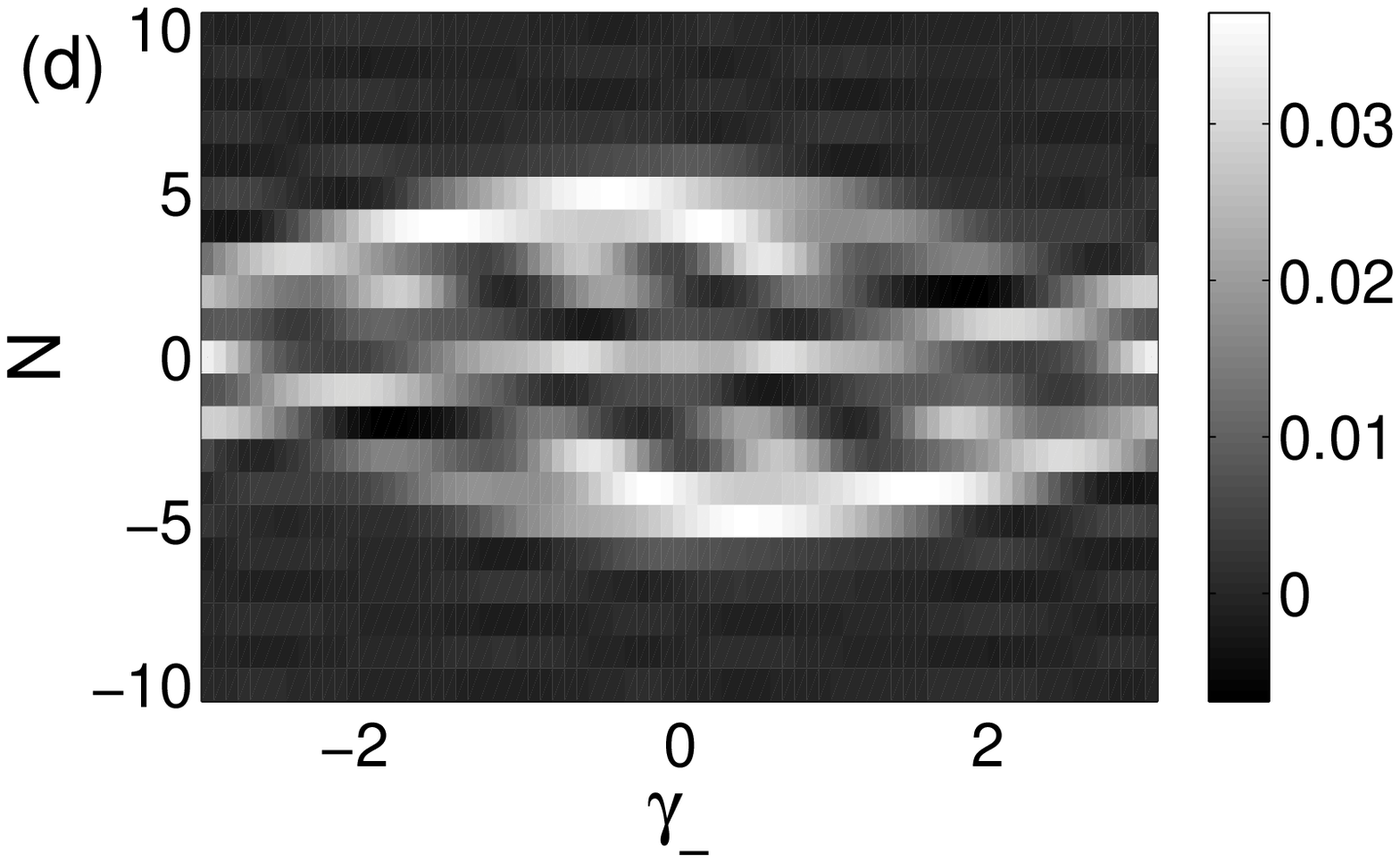}}
 \caption{Spreading and interference during the early stages of quantum evolution. Wigner functions for a system where $\omega_d$ is switched from $0$ to $1$ at $t=0$: (a) $\tau/\tau_d=0$, (b) $\tau/\tau_d=0.477$, (c) $\tau/\tau_d=0.716$, (d) $\tau/\tau_d=0.955$. The gate resistance is $50\Omega$, with other parameters  $E_J=\cos(0.2)$, $E_{C_J}=\sin(0.2)$, $N_g=0$, and $k_{\rm B}T_g=2E_J$. The initial state is the steady-state of the system with $\omega_d=0$.}\label{fig:qcoh}
\end{figure}

We start by considering how the system evolves starting from an initial state which is `classical' in the sense that its Wigner function is  relatively smooth as a function of $N$. The initial state we choose to use (illustrated in Fig.\ \ref{fig:qcoh}a) is the steady-state of the system when the dc voltage is switched off (i.e.\ the $\omega_d=0$ limit). and hence should be easy to prepare in practice. Figure \ref{fig:qcoh} shows a series of snapshots of the Wigner function at different times after we set $\omega_d=1$ at time $t=0$. The Wigner function  becomes stretched, reaches the edge of the phase space, and then starts to warp around on itself. This wrapping around leads to interference and the formation of regions where the Wigner function is negative. Very similar results are obtained when the system is initially in a coherent state instead.

\begin{figure}
\center{
\includegraphics[width=6.5cm]{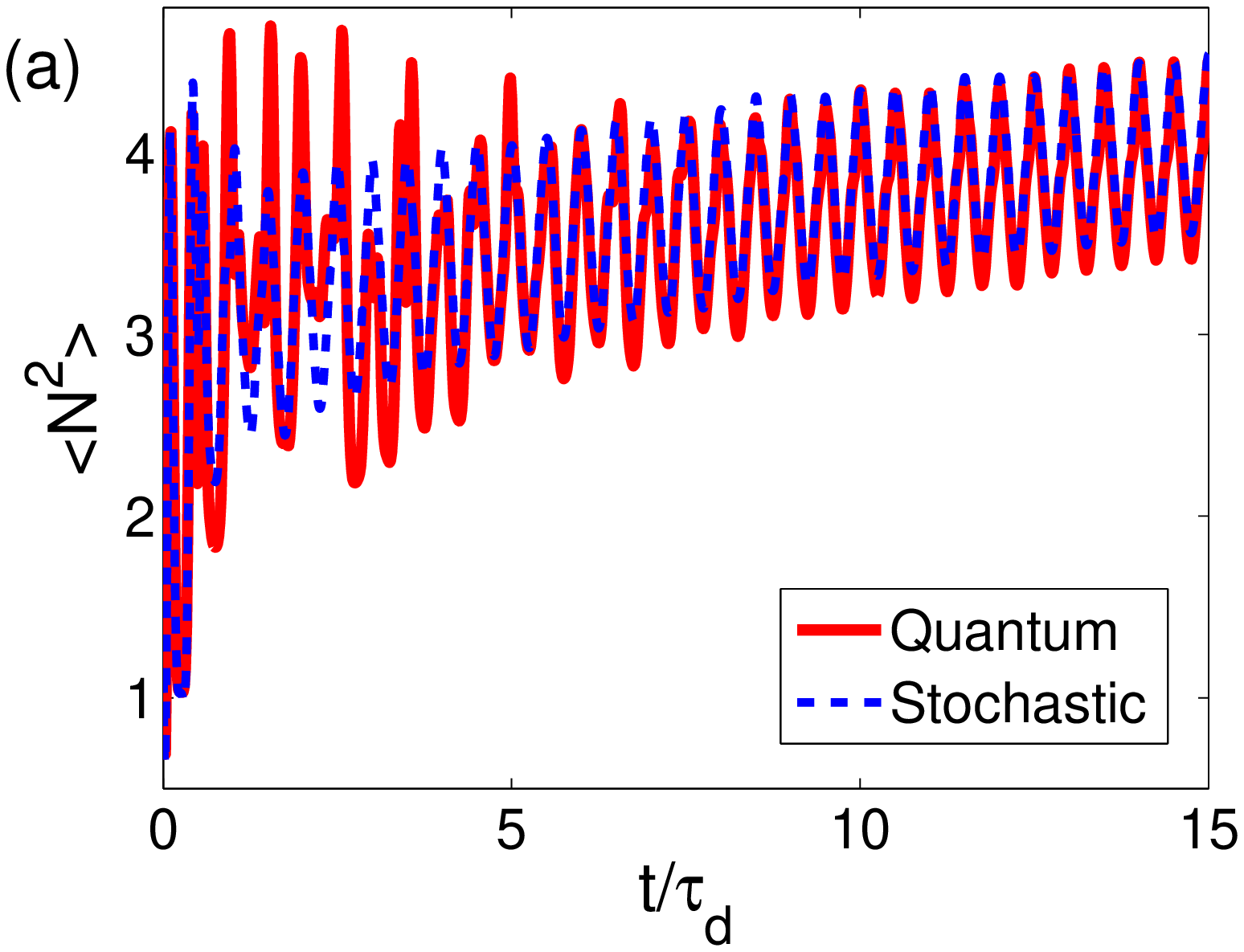}
\includegraphics[width=6.5cm]{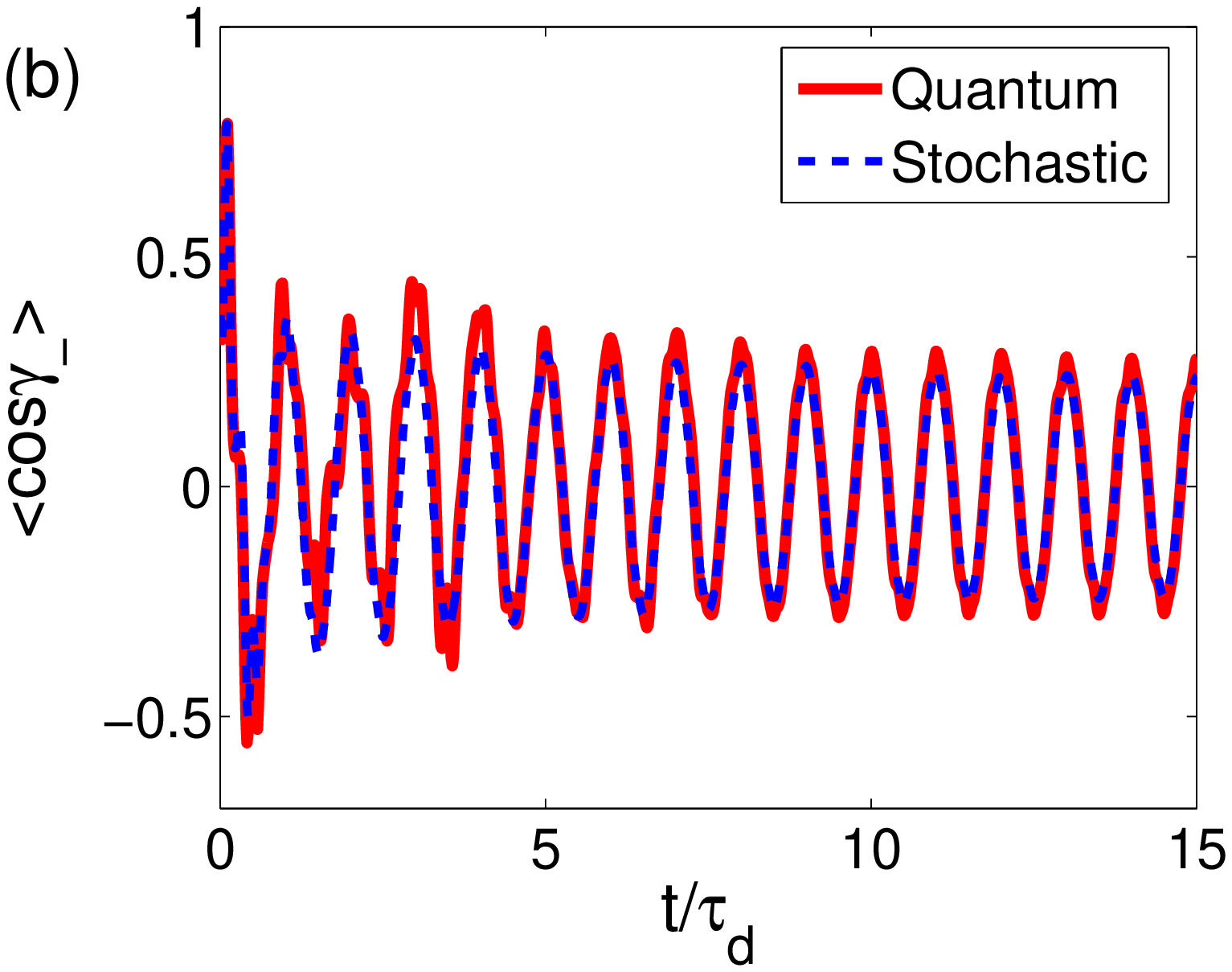}}
 \caption{Comparison of the dynamics of the stochastic and quantum evolution (a) $\langle N^2\rangle$ (b) $\langle \cos(\gamma_-)\rangle$. The parameters chosen are: $E_J=4$, $E_{C_J}=1$, $N_g=0$, $k_{\rm B}T_g=2E_J$ and $R_g=500\Omega$. The initial state for the quantum (stochastic) dynamics was a coherent state (Gaussian distribution) centered on the point $N=\gamma_-=1$.}\label{fig:moments}
\end{figure}

\begin{figure}
\center{
\includegraphics[width=6.5cm]{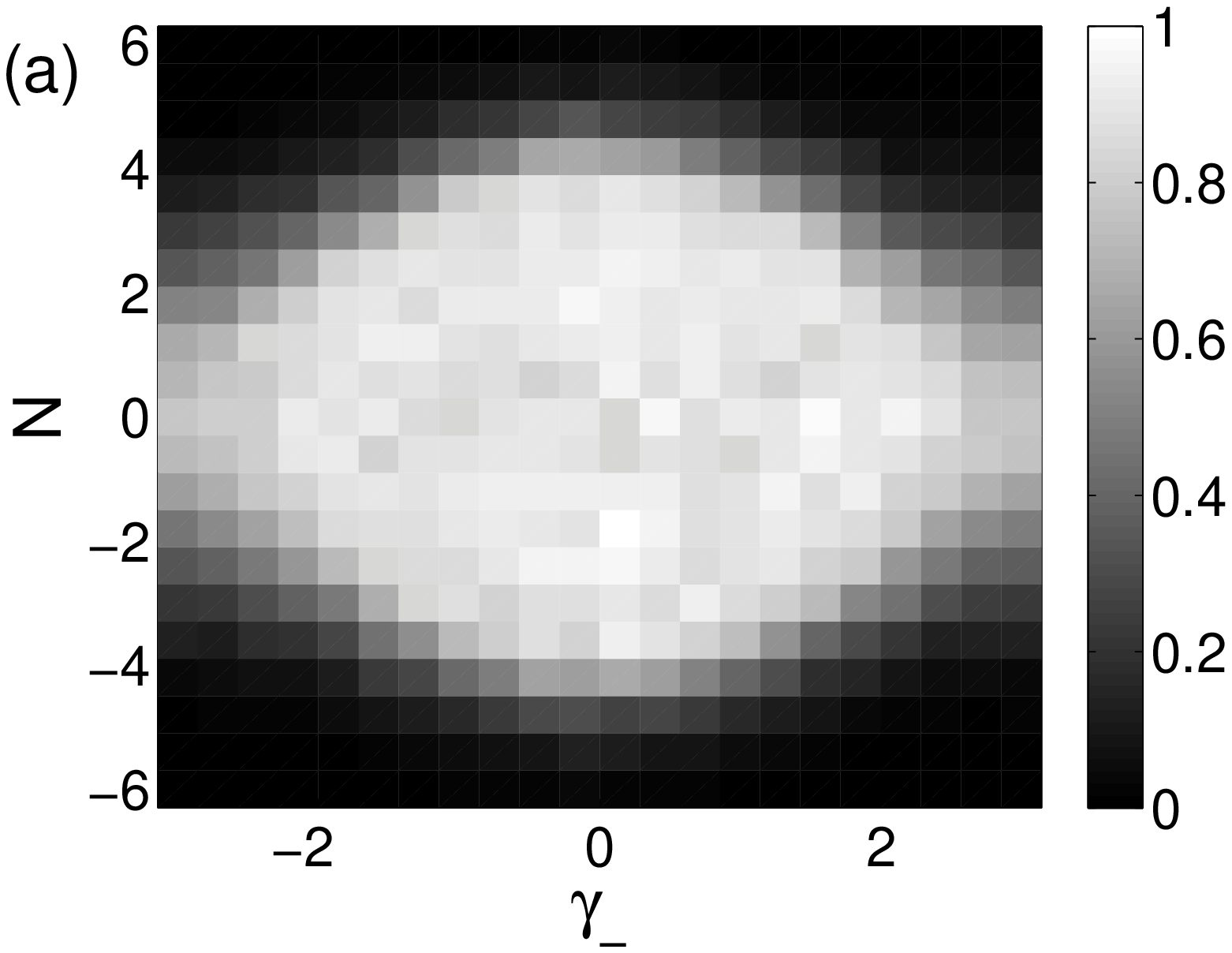}
\includegraphics[width=6.5cm]{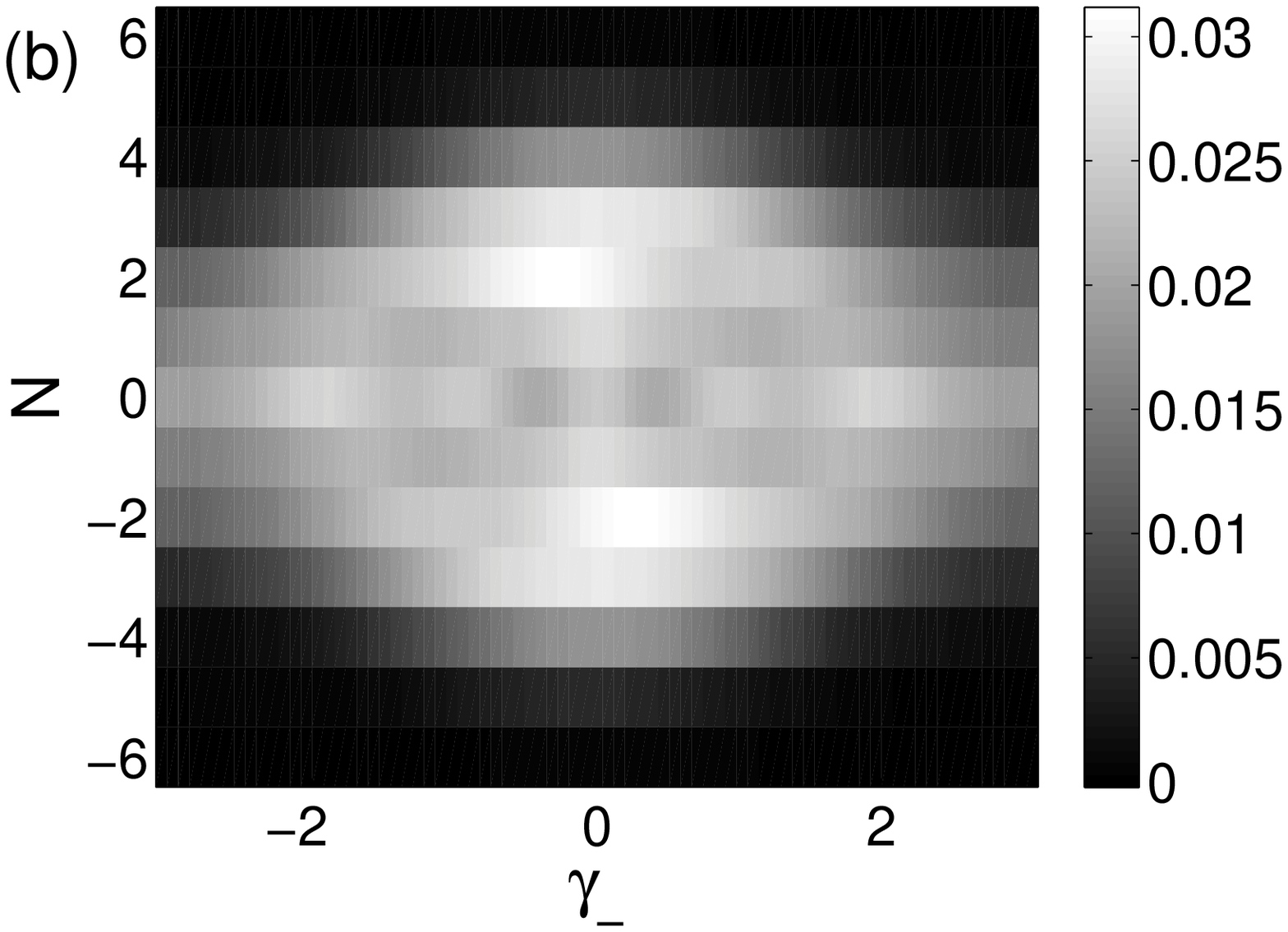}
}
 \caption{(a) Stochastic probability distribution and (b) Wigner distribution of the corresponding quantum dynamics at $\tau=75\tau_d$. The parameters are the same as in Fig.\ \ref{fig:moments}.}\label{fig:qdist}
\end{figure}

Over longer time-scales and for sufficiently large $E_J/E_{C_J}$ the quantum dynamics matches up fairly well with the stochastic dynamics (described in Sec.\ \ref{sec:classd}). A good overview of the dynamics can be obtained by looking at some of the moments as a function of time, as shown in Fig.\ \ref{fig:moments}. The behavior of $\langle N^2\rangle$ and $\langle\cos\gamma_-\rangle$ in both the quantum and stochastic dynamics is eventually periodic, a consequence of the underlying periodicity of the drive. The average charge, $\langle N\rangle$ decays rapidly to zero in both cases. The classical stochastic dynamics comes close to reproducing the behavior seen in the quantum dynamics of both $\langle N^2\rangle$ and $\langle\cos\gamma_-\rangle$ in the long time limit, although the amplitude of the stochastic oscillations is slightly smaller than the quantum ones. Going beyond the moments of the system, we can compare the full stochastic probability distribution with the Wigner function of the corresponding quantum evolution as shown in Fig.\ \ref{fig:qdist}.

\begin{figure}
\center{
\includegraphics[width=6.5cm]{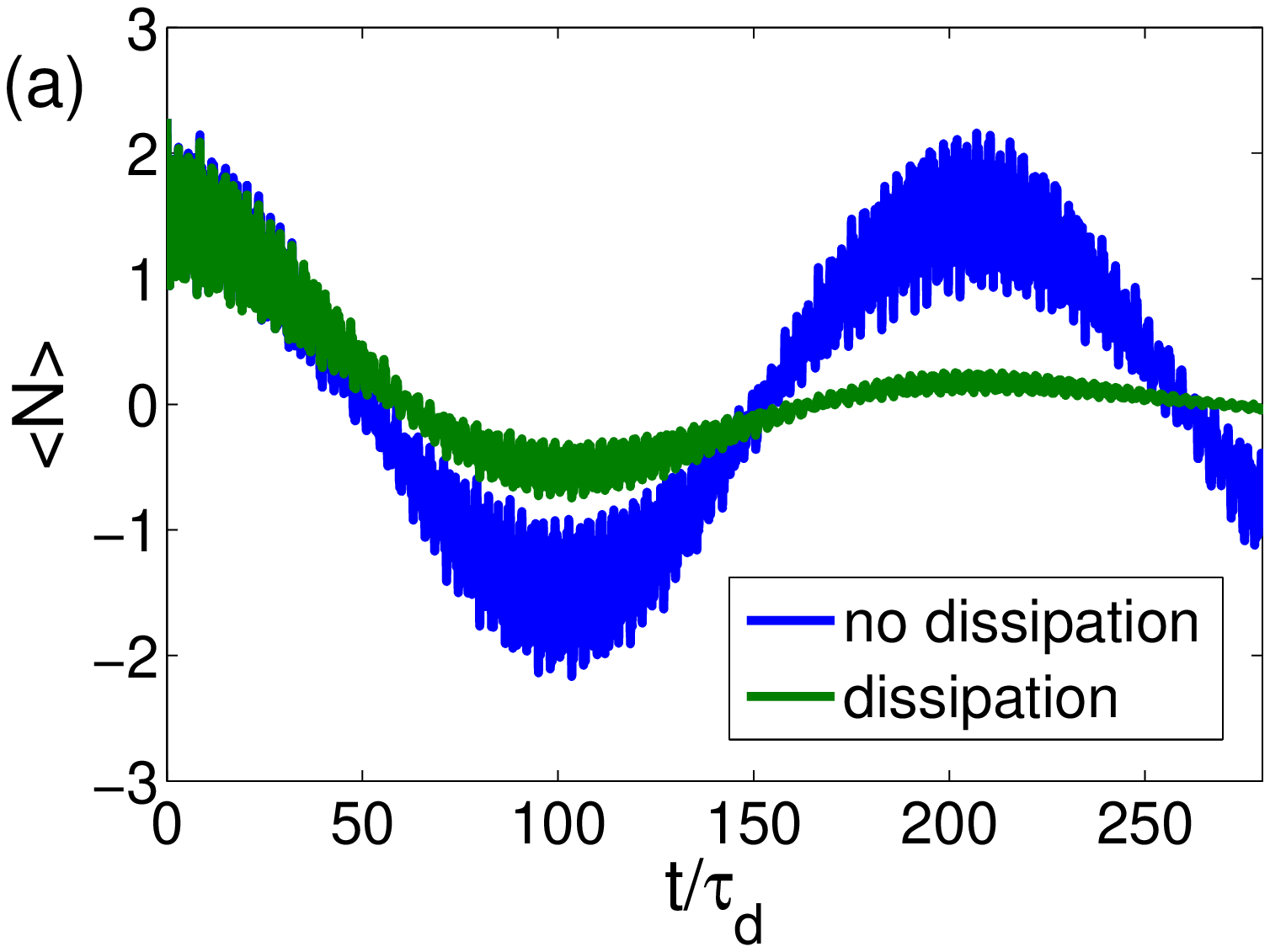}
\includegraphics[width=6.5cm]{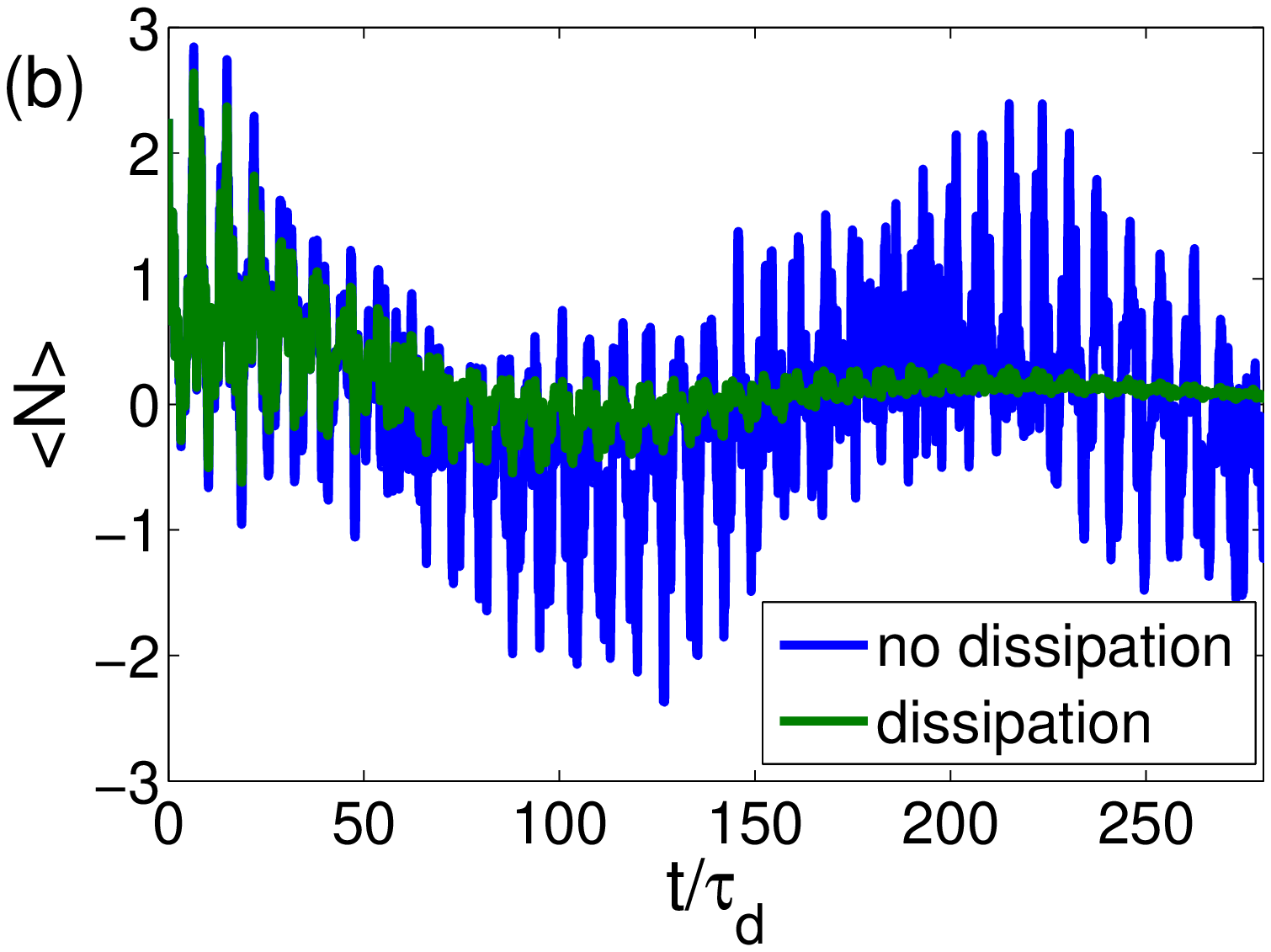}
}
 \caption{Comparison of the evolution of $\langle N\rangle$ with ($R_g=50\Omega$) and without damping for an initial coherent state centered on a point (a) inside an island of the classical phase space $(N,\gamma_-)=(2.27,0)$ (b) inside the chaotic sea $(N,\gamma_-)=(2.27,1.25)$. The parameters used match those in Fig.\ \ref{fig:class1}b.}\label{fig:qtun}
\end{figure}

\begin{figure}
\center{
\includegraphics[width=6.5cm]{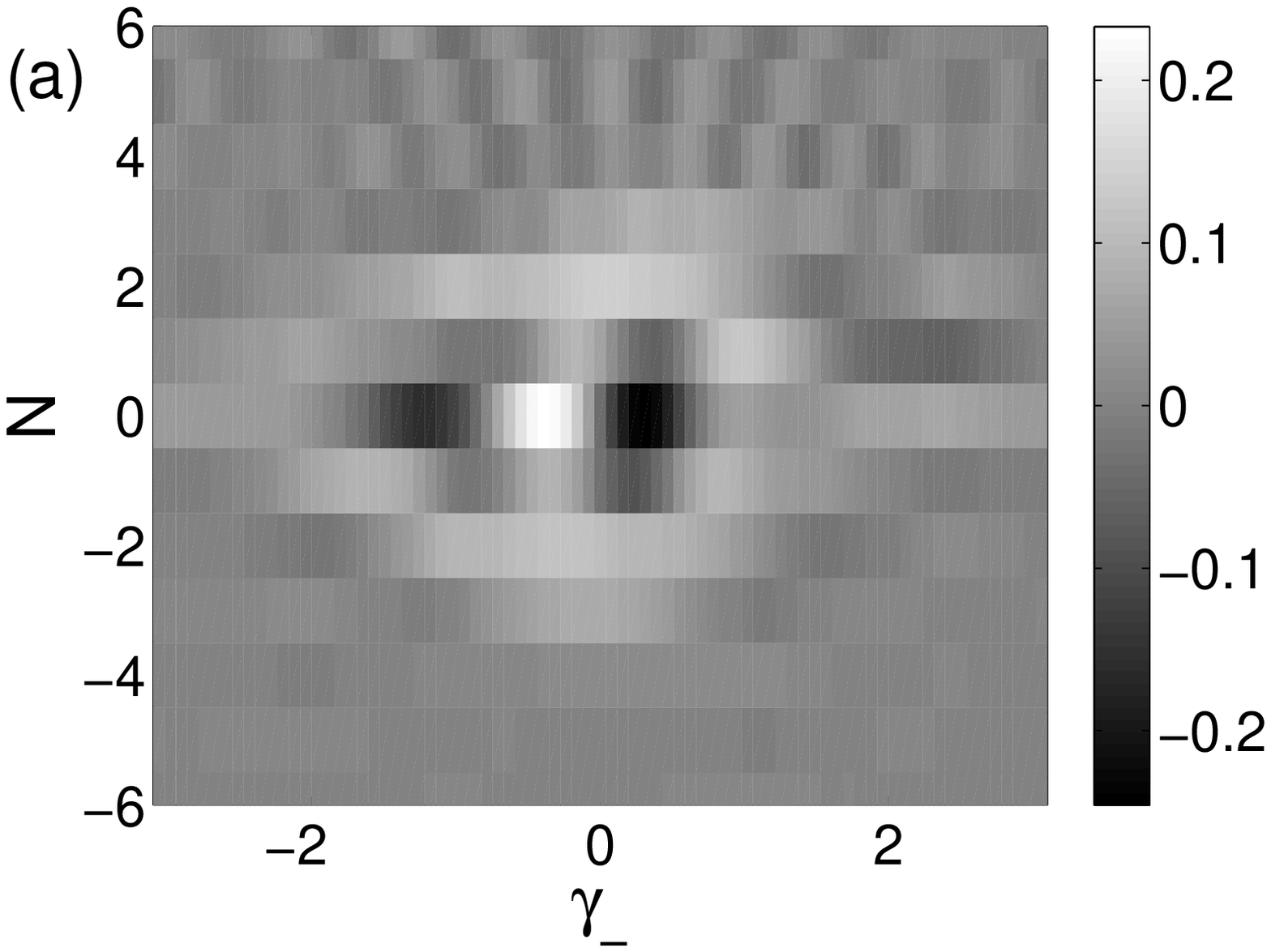}
\includegraphics[width=6.5cm]{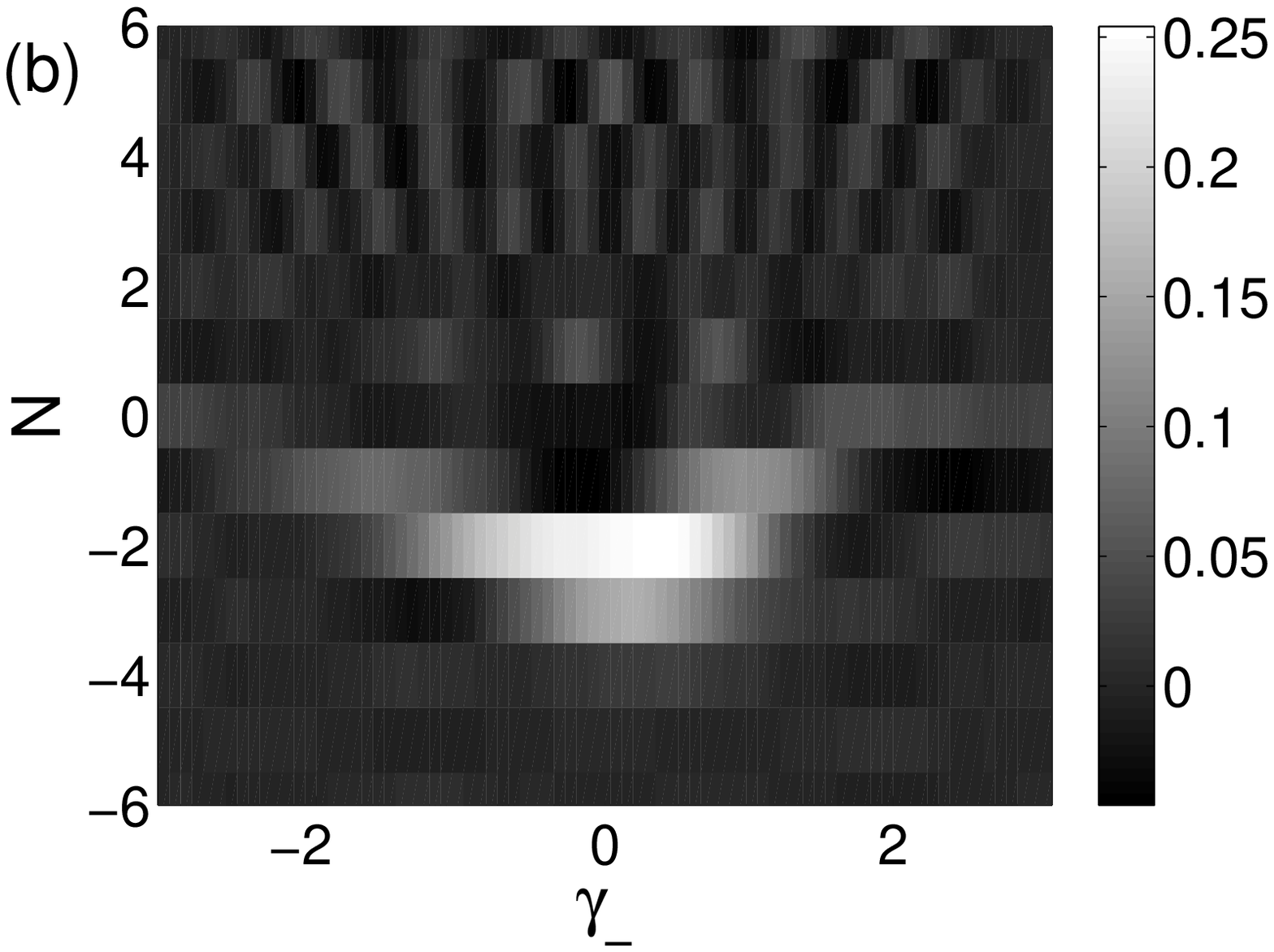}
}
 \caption{Wigner functions at (a) $\tau=50\tau_d$ and (b) $\tau=100\tau_d$ for the parameters corresponding to Fig.\ \ref{fig:qtun}a with $R_g=0$.}\label{fig:qtunw}
\end{figure}

Different choices of initial conditions corresponding to different regions of the classical phase space lead to rather different quantum evolutions, although again this is only a transient effect in the presence of dissipation. Analyses of similar systems~\cite{mouchetpre01,mouchetpre06} (without dissipation) have shown that they can display chaos-assisted quantum tunneling~\cite{davisjcp81,ballentine90,tomsovic94,latkapra94,kohlerpre98,steckscience01,mouchetpre01,mouchetpre06,chaudhurynature09}. In the classical dynamics any point initially chosen to be within one of the stable islands is confined to either the $N>0$ or $N<0$ part of the phase space (depending on the initial conditions) and the value of $N$ oscillates quasi-periodically on a timescale $\tau_d/2$. In the quantum dynamics, the system can tunnel between these two regions of phase space. We looked at the evolution of an initial coherent state centered on points which are either within a stable island or the chaotic sea in the classical phase space (with parameters which match those of Fig.\ \ref{fig:class1}b).
The dynamics of the average charge for these two cases is compared in Fig.\ \ref{fig:qtun}. An initial state centred on a stable island  (Fig.\ \ref{fig:qtun}a) shows fast oscillations in $\langle N\rangle$, which have the same time-scale and (initially) a similar amplitude to those seen in the classical dynamics. However, there is also an underlying much slower oscillation which takes the system to the opposite side of the phase space. In contrast, for an initial coherent state centered in the chaotic sea (Fig.\ \ref{fig:qtun}b) the behavior is much less regular. The differences in behavior wash out very rapidly when dissipation is included, an important sign that the slow oscillations involve coherent superpositions. The slow oscillations in Fig.\ \ref{fig:qtun}a  take the system from a state localized around one of the regions of phase space corresponding to a classical island of stability, to one which is largely localized on the other stable island. The system tunnels via a state which shows strong interference effects, as can be seen from the Wigner functions calculated after a quarter and after a half of the slow oscillation period shown in Fig.\ \ref{fig:qtunw}.

\section{Classical limit}
\label{classicallimitsec}

A  Wigner function's domain is given by the eigenvalue spectra of the  canonically conjugate operators from which the function is constructed. For the number/phase Wigner function (see Sec.~\ref{wignersec}), the domain consists of parallel lines  [$\gamma_-=(-\pi,\pi)$ and $N=0, \pm1, \pm2,\dots$].  Furthermore, since  these eigenvalues are the possible outcomes of measurements of the associated observables, it is natural to use the Wigner function to define a quantum phase space dynamics from which the classical phase space limiting dynamics can be recovered by approximation. In the following, we again restrict ourselves to the driven pendulum subsystem dynamics only, resulting from formally setting  $\Delta^+_{zp}= 0$ in the Hamiltonian~(\ref{quantumhamiltonianeq}).

In contrast to a nonlinear system with the configuration space topology $\mathbb{R}$, such as the commonly investigated one-dimensional Duffing oscillator, it does not appear possible to derive a closed form  equation for the Wigner function $W_N(\gamma_-,t)$ starting from the master equation~(\ref{markovgatemastereq}). The  problem lies in the $\sin\gamma_-$ potential and damping  terms, which give rise formally to Wigner functions shifted by $1/2$ in their $N$-arguments. Because of the non-trivial $S^1$ configuration space topology,  the $N$ coordinate domain of the Wigner function consists of the discrete integers, not half-integers. On the other hand, it is possible to write down closed form equations for the definite parity functions $w_{N+\mu/2}(\gamma_-,t)$. From~(\ref{markovgatemastereq}) and (\ref{parityfunctioneq}), we obtain after  some algebra:
\begin{eqnarray}
&&\frac{\partial w_{N+\mu/2}}{\partial t}=-\frac{2 E_{C_J}}{\hbar} (N+\mu/2 -N_g) \frac{\partial w_{N+\mu/2}}{\partial \gamma_-}\cr
&&-\frac{2 E_J}{\hbar}\sin\gamma_- \cos (\omega_d t)\left(w_{N-1/2 +\mu/2}-w_{N+1/2 +\mu/2}\right)\cr
&&+\frac{1}{\hbar^2}\left(\frac{e C_g}{C_J}\right)^2 k_B T_g R_g \frac{\partial^2 w_{N+\mu/2}}{\partial \gamma_-^2}\cr
&&-\frac{1}{\hbar^2}\left(\frac{e C_g}{C_J}\right)^2 E_J R_g \frac{\partial}{\partial\gamma_-}\left[\sin\gamma_- \cos(\omega_d t)\left(w_{N-1/2 +\mu/2}+w_{N+1/2 +\mu/2}\right)\right].\cr
&&\label{paritywignermastereq}
\end{eqnarray}
Using the $\gamma_-$ integral form of the Wigner function in (\ref{circlewignereq}) to analytically continue the function to arbitrary real $N$, we obtain from Eqs.~(\ref{paritywignermastereq}) and (\ref{circlewignereq}) the following equation for the Wigner function:
\begin{eqnarray}
\frac{\partial W_{N}}{\partial t}&=&-\frac{2 E_{C_J}}{\hbar} (N -N_g) \frac{\partial W_{N}}{\partial \gamma_-}\cr
&& -\frac{2 E_J}{\hbar}\sin\gamma_- \cos (\omega_d t)\left(W_{N-1/2 }-W_{N+1/2}\right)\cr
&&+\frac{1}{\hbar^2}\left(\frac{e C_g}{C_J}\right)^2 k_B T_g R_g \frac{\partial^2 W_{N}}{\partial \gamma_-^2}\cr
&&-\frac{1}{\hbar^2}\left(\frac{e C_g}{C_J}\right)^2 E_J R_g \frac{\partial}{\partial\gamma_-}\left[\sin\gamma_- \cos(\omega_d t)\left(W_{N-1/2}+W_{N+1/2}\right)\right].\cr
&&\label{wignermastereq}
\end{eqnarray}
Under conditions where $W$ varies slowly with $N$,  we can Taylor expand the fractionally-shifted $W$ functions to first order as a good approximation and obtain the following classical master equation for the probability distribution $P(\gamma_-,N,t)$:
\begin{eqnarray}
\frac{\partial P}{\partial t}&=&-\frac{2 E_{C_J}}{\hbar} (N -N_g) \frac{\partial P}{\partial \gamma_-}+\frac{2 E_J}{\hbar}\sin\gamma_- \cos (\omega_d t)\frac{\partial P}{\partial N}\cr
&&+\frac{1}{\hbar^2}\left(\frac{e C_g}{C_J}\right)^2 k_B T_g R_g \frac{\partial^2 P}{\partial \gamma_-^2}\cr
&&-\frac{2}{\hbar^2}\left(\frac{e C_g}{C_J}\right)^2 E_J R_g \frac{\partial}{\partial\gamma_-}\left[\sin\gamma_- \cos(\omega_d t)P\right].
\label{classicalmastereq}
\end{eqnarray}
This master equation is equivalent to the classical pendulum  Langevin equation~(\ref{dimlesslangevineq}), with $(\gamma_+,p_+)$ set to zero.

In contrast to the usual situation for a nonlinear system with trivial configuration space topology $\mathbb{R}$, the classical limit  (\ref{classicalmastereq}) of the quantum master  equations (\ref{paritywignermastereq}) and (\ref{wignermastereq}) was not obtained by identifying and then discarding a higher derivative quantum term involving the anharmonic system potential~\cite{zurekprl94}. Rather, the difference between the quantum and classical equations is more subtle and  linked to the discreteness of the number (equivalently angular momentum) operator, which in turn is a consequence of the non-trivial configuration space topology.  All that is required to recover the classical master equation to a good approximation is that the Wigner function varies by only a small amount as  its argument $N$ increases or decreases  by one.

Clearly, the pendulum state must have non-negligible overlap with a large number of angular momentum eigenstates $|N\rangle$  if the Wigner function is to depend smoothly on $N$; recovering the pendulum classical limit necessarily requires  $E_{C_J}\ll E_J$, e.g., a `transmon'-like CPT~\cite{kochpra07}. However, the latter inequality is not a sufficient condition:  as shown in Fig.~\ref{fig:qcoh}, an initially `classical', i.e., smooth, positive practically everywhere Wigner function can  evolve through stretching and shrinking into a Wigner function that is non-smooth in $N$, so that the classical master equation approximation~(\ref{classicalmastereq})  breaks down.  Furthermore, because of the $S^1$  periodicity of the configuration space, the stretching pendulum wavefunction can eventually interfere with itself, resulting in an oscillatory Wigner function with significant negative regions.

Even though  the initial state-dependent `transient' dynamics will display quantum features, the dynamics will eventually settle into a steady state where the Wigner function is practically positive everywhere and  well-approximated by the classical master equation~(\ref{classicalmastereq}), provided $k_BT_g\gtrsim E_J\gg E_{C_J}$.  The smaller is the gate voltage resistance $R_g$ (or gate capacitance $C_g$),  the longer is the duration of the transient quantum interval.

Interestingly,  when the charging energy is not small, i.e., $E_{C_J}\gtrsim E_J$, the steady state Wigner distribution will  still be practically positive everywhere and hence interpretable as a probability density, provided $k_BT_g\gtrsim E_{C_J}$. However, because the Wigner function is non-smooth and  with non-negligible support over only a small range in $N$, the classical pendulum master equation no longer accurately describes the Wigner function dynamics. The question then arises as to whether there is an approximate classical description that is distinct from the classical pendulum equation. Such an equation must necessarily treat $N$ as a discrete coordinate and so is more appropriately interpreted in terms of the charge dynamics of the CPT.  For   $E_J\ll E_{C_J}$ and sufficiently large gate voltage resistance and effective noise temperature, $P(E)$ theory may provide an adequate classical description in terms  of incoherently tunneling Cooper pairs~\cite{ingold92,ingoldprb94,leppakangasprb06}. Otherwise, Cooper pair tunneling across the Josephson junction is an inherently quantum coherent process: even though the steady state Wigner function evolution is practically positive everywhere, the model dynamics must be  interpreted as quantum in nature when the charging energy is large.

\section{Conclusion}
\label{sec:conclusion}
In this chapter, we have investigated a strongly non-linear superconducting device consisting of a Cooper pair transistor (CPT) that is coupled to a dc voltage biased microwave cavity and driven by the dc  bias via the ac Josephson effect. Our main focus has been on comparing the quantum and classical dynamics of the system -- the "quantum-classical correspondence" -- in particular establishing the circumstances under which the corresponding dynamics are similar.

We  derived the corresponding classical Langevin and quantum master equations of motion, which describe the dynamics of the CPT-cavity system in the presence of an environment consisting of dissipative circuit elements. Although we did not investigate the dynamics of the full system, we did explore the dynamics of the driven CPT (which is the non-linear element of the device) in the limit where the cavity does not affect its behavior.  The quantum-classical correspondence was elucidated by considering the Wigner function representation of the evolving CPT quantum state.  Numerical simulation of the quantum and classical dynamics shows that, apart from initial state-dependent transients, the evolution becomes very similar in the limit of large Josephson energies, for which  the discreteness of the Cooper pair number on the CPT island is unimportant. Interestingly, though, the transient behavior of the system can lead to highly non-classical states even when the initial state is apparently very classical.

Future work will need to explore how the full system (CPT and cavity degrees of freedom) behaves, as well as connect the dynamics of the system to quantities that are measured in experiment.  On a more technical level, further analysis of the interaction between the CPT degrees of freedom and the gate impedance is needed in order to derive the correct description of the dissipative dynamics in the low temperature limit.

\section*{Acknowledgements}
M.P.B. thanks F. Nori and the Advanced Science Institute, RIKEN, for their hospitality and support, where part of this work was carried out.  MPB acknowledges support from the NSF under grant number DMR-0804477, ADA acknowledges support from the EPSRC  under grant number  EP/E034442X/1, and AJR is supported by  the NSF and AFOSR/DARPA under grant numbers DMR-0804488 and FA9550-10-1-0047, respectively.

\end{document}